\documentclass[a4paper,fleqn,usenatbib,referee]{mnras}

\usepackage[dvipdfmx]{graphicx}
\usepackage{amsmath}	
\usepackage{amssymb}	
\usepackage{bm}
\usepackage{multirow}
\usepackage{ulem} 

\newcommand{\msun}{\thinspace M_\odot} 
\newcommand{\vect}[1]{\mbox{\boldmath$#1$}}
\newcommand{\cm}{\,{\rm cm}^{-3} }

\title[Close Binary Formation]
{Impact of Magnetic Braking on High-mass Close Binary Formation }

\author[Harada et al.]{
Naoto Harada$^{1}$\thanks{E-mail: harada.naoto.450@s.kyushu-u.ac.jp (KTS)},
Shingo Hirano$^{1,2}$,
Masahiro N. Machida$^{1}$,
and Takashi Hosokawa$^{3}$
\\
$^{1}$Department of Earth and Planetary Sciences, Faculty of Sciences,
	Kyushu University, Fukuoka 819-0395, Japan\\
$^{2}$Department of Astronomy, School of Science, University of Tokyo, Bunkyo, Tokyo 113-0033, Japan\\
$^{3}$Department of Physics, Kyoto University, Kyoto 606-8502, Japan
}

\date{Accepted XXX. Received YYY; in original form ZZZ}

\pubyear{2015}

\begin{document}
\label{firstpage}
\pagerange{\pageref{firstpage}--\pageref{lastpage}}
\maketitle

\begin{abstract}
Combining numerical simulations and analytical modeling, we investigate whether close binary systems form by the effect of magnetic braking.
Using magnetohydrodynamics simulations, we calculate the cloud evolution with a sink, for which we do not resolve the binary system or binary orbital motion to realize long-term time integration.
Then, we analytically estimate the binary separation using the accreted mass and angular momentum obtained from the simulation.
In unmagnetized clouds, wide binary systems with separations of $> 100$\,au form, in which the binary separation continues to increase during the main accretion phase. 
In contrast, close binary systems with separations of $<100$\,au can form in magnetized clouds.
Since the efficiency of magnetic braking strongly depends on both the strength and configuration of the magnetic field, they also affect the formation conditions of a close binary.
In addition, the protostellar outflow has a negative impact on close binary formation, especially when the rotation axis of the prestellar cloud is aligned with the global magnetic field. 
The outflow interrupts the accretion of gas with small angular momentum, which is expelled from the cloud, while gas with large angular momentum preferentially falls from the side of the outflow onto the binary system and widens the binary separation.  
This study shows that a cloud with a magnetic field that is not parallel to the rotation axis is a favorable environment for the formation of close binary systems.
\end{abstract}




\section{Introduction}
\label{sec:intro}
Recent observations have shown that most high-mass stars are in close binary systems.
The mass ratio of the primary to the secondary stars $q$ in such systems is typically $q\sim0.5$, while there exists a non-negligible fraction of massive twins $q>0.95$ \citep{Duchene2013,moe17}.
Thus, it is expected that high-mass stars tend to be born as close binary systems, in which some systems have nearly equal mass.
High-mass close binaries might be the origin of binary black holes or binary neutron stars detected  by gravitational waves \citep{abbott21}.
Therefore, it is crucial to understand the formation process of such binaries. 

When investigating the formation of (high-mass) binary systems, the most important factor is the angular momentum \citep{Lund2018}.
The binary separation is determined by the angular momentum of  the system. 
In other words, the binary formation  is related to  how much angular momentum is introduced into the system and how much is removed or transported from it.
In the star formation process, angular momentum is naturally introduced into the system by mass accretion during the main accretion phase, because the infalling gas possesses angular momentum.  
It can be  expected that the separation of the protobinary system will gradually widen over time, because parcels of gas with large angular momentum will  later accrete onto the system, as shown in \citet{satsuka2017}.
However, many observations of (high-mass) close binary systems in various star-forming regions \citep{Duchene2013} indicate that there is likely to be  a mechanism to efficiently remove angular momentum from the system in the binary formation process. 
It is considered that the most efficient mechanism to remove angular momentum from the system is magnetic braking \citep{mestel56,gillis79,mouschovias79}. 
Recent observations also imply that the magnetic field in high-mass star-forming cores is very strong \citep{li15}. 
If a (strong) magnetic field is considered in the binary formation process,  the angular momentum of the binary system should be  transported into the infalling envelope due to magnetic braking.
Since each star of a protobinary system is  connected to the massive infalling envelope through the magnetic field \citep{machida09},  a massive envelope can brake the binary orbital motion and shrink the binary separation.
To date,  magnetic braking has been mainly investigated for the single star formation process \citep{zhao20}.
However,  the effect of  the magnetic field and magnetic braking should also be considered for the binary formation process. 

In the framework of the core collapse scenario, fragmentation occurs in the star forming cloud and a binary or multiple system forms \citep{tsuribe99, cha03,matsumoto03}.
Without considering the magnetic field, the formation and evolution of binary systems composed of low-mass stars have been investigated in many studies \citep{Bonnell1994, bate1997, bate2009a, bate2009c, bate2012, BateBonnell1997, bate2005, bate2002}.
Recently, the formation of a high-mass binary system in unmagnetized clouds are also investigated \citep{bonnell2005, meyer18}.
On the other hand, while the effect of the magnetic field on the fragmentation process has been investigated \citep[e.g.][]{machida08,hennebelle08}, 
there are few studies investigating the binary orbital motion or separation during the main accretion phase after fragmentation  in a magnetized cloud.
Using three-dimensional (3D) magnetohydrodynamics (MHD) simulations, \citet{Kuruwita2017} investigated binary formation, in which they calculated the evolution of a binary system for $\sim3000$\,yr after fragmentation.
They showed that angular momentum is efficiently transferred due to  magnetic braking, and the binary system maintains a separation within $\sim 5$\,au, in which a sink radius of $\sim4.8$\,au was adopted to accelerate the calculation. 
Thus, they could not precisely resolve close binary systems within a separation of $\lesssim 5$\,au.
\citet{Saiki2020} also investigated close binary formation in a resistive MHD simulation and  found that the magnetic field plays a significant role in shrinking the binary separation during  the main accretion phase.
However, since  the protostars in their study were resolved with a spatial resolution of $0.039$\,au, they calculated the evolution of a proto-binary system only for $\sim400$\,yr after the protostars formed. 
Although 3D MHD simulations are appropriate for investigating binary formation, it is difficult to calculate and investigate the long-term evolution of protobinary systems. 
The binary orbital motion (or binary orbital period) should be resolved in such simulations. 
However, the binary orbital period shortens as the binary separation shrinks. 
Therefore, 3D simulations are not suitable for investigating the long-term evolution of  close binary systems.

Recently,  \citet{Lund2018} investigated the formation of high-mass close binaries with a new approach, in which they analytically estimated how much angular momentum is  transported by magnetic braking, and concluded that high-mass close binaries with separations of $\lesssim10$--$100$\,au can form when the prestellar cloud is strongly magnetized. 
However, since the configuration and strength of the magnetic field are idealized in their study, it is difficult to correctly estimate the efficiency of  magnetic braking.
\citet{Hirano2020} showed that the efficiency of  magnetic braking changes significantly over time during the main accretion phase, because the configuration of  the magnetic field varies over time. 
Thus, to investigate the effect of magnetic  braking on the binary separation, a realistic treatment of the magnetic field (such as its configuration and strength) is necessary. 

In studying close binary formation, details of the long-term evolution and a realistic treatment of  the magnetic field are both required.
However, as seen in \citet{Saiki2020}, it is not possible to calculate the long-term evolution of a binary system while resolving the orbital motion. 
In analytical studies, it is difficult to precisely estimate the efficiency of magnetic braking and the angular momentum introduced into the binary system, which determines the binary separation, without knowing the correct configuration and strength of the magnetic field.  

In this study, we calculate the configuration and strength of  the magnetic field using 3D MHD simulations with a sink, in which we do not resolve the binary orbital motion to realize long-term time integration. 
Then, using data obtained  from the simulation (the mass and angular momentum falling onto the sink), we analytically estimate the binary separation according to the prescription used in \citet{Lund2018}.
This paper is structured as follows. 
\S2 describes the analytical method and numerical settings of our model. The results are presented in \S3. 
We discuss the relation between the density distribution and angular momentum transfer and the dependence of the results on the spatial resolution in \S4. 
Our conclusions are presented in \S5.

\section{Methods}
\label{sec:methods}
The purpose of this study is to investigate the evolution of binary systems during the main accretion phase taking into account magnetic effects such as magnetic braking and protostellar outflow. 
According to  \citet{Lund2018}, we estimate the binary parameters analytically, while we use the data obtained from MHD simulations to estimate them. 
In this section, we first explain how to analytically estimate the binary parameters
and then describe the settings of our MHD simulations.

\subsection{Binary Separation and Accretion of Mass and Angular Momentum}
\label{sec:separation}
Assuming a circular orbit, the binary orbital angular momentum $J_{\rm b}$ is related to the total stellar mass $M_{\rm tot}$ and binary separation $r_{\rm sep}$ as  
\begin{equation}
J_{\rm b} = \frac{q}{(1+q)^2} \sqrt{G M_{\rm tot}^3\,r_{\rm sep}},
\label{eq:angmom}
\end{equation}
where $q$ is the mass ratio of the primary to the secondary stars \citep[e.g.][]{BateBonnell1997}.
Since this study focuses on the binary separation,  equation~(\ref{eq:angmom}) is transformed to 
\begin{equation}
r_{\rm sep} = \frac{(1+q)^4}{q^2} \frac{J_{\rm b}^2}{G M_{\rm tot}^3}.
\label{eq:separation1}
\end{equation}
Equation~(\ref{eq:separation1}) indicates that determining $r_{\rm sep}$ requires $J_{\rm b}$, $M_{\rm tot}$, and $q$.
During the main accretion phase, mass and angular momentum are continuously supplied from the infalling envelope into the binary system. 
Thus, $M_{\rm tot}$ and $J_{\rm b}$ of the binary system vary over time. 
For simplicity, we assume an equal mass binary  ($q=1$) in the following. 
Then, equation~(\ref{eq:separation1}) is described by
\begin{equation}
r_{\rm sep} = \frac{16}{G} \frac{J_{\rm b}^2}{M_{\rm tot}^3}
= \frac{16}{G} \frac{j_{\rm b}^2}{M_{\rm tot}},
\label{eq:separation2}
\end{equation}
where $j_{\rm b}(\equiv J_{\rm b}/M_{\rm tot})$ is the specific angular momentum of the binary system. 
$r_{\rm sep}$ varies during the mass accretion phase, because both $M_{\rm tot}$ and $J_{\rm b}$ gradually increase due to  the mass accretion. 
Here, we define the angular momentum and mass of the accreting matter as $\Delta J_{\rm acc}$ and $\Delta M_{\rm acc}$, respectively.
When $\Delta J_{\rm acc}$ and $\Delta M_{\rm acc}$ are introduced into the protobinary system by accretion, $r_{\rm sep}$ changes  as
\begin{equation}
    r_{\rm sep} + \Delta r_{\rm sep}
    = \frac{16}{G}\frac{(J_{\rm b}+\Delta J_{\rm acc})^2}{(M_{\rm tot}+\Delta M_{\rm acc})^3},
\label{eq:separation3}
\end{equation}
where $\Delta r_{\rm sep}$ is the change in the binary separation due to the accretion of mass and angular momentum, described by
\begin{align}
    \Delta r_{\rm sep}
    &= \frac{16}{G}\left( \frac{(J_{\rm b}+\Delta J_{\rm acc})^2}{(M_{\rm tot}+\Delta M_{\rm acc})^3}
    - \frac{J_{\rm b}^2}{M_{\rm tot}^3} \right) \label{eq:delta_s0}  \\
    &\sim \frac{16}{G}\frac{J_{\rm b} (2M_{\rm tot} \Delta J_{\rm acc}-3J_{\rm b} \Delta M_{\rm acc})}{M_{\rm tot} (M_{\rm tot}^3+3M_{\rm tot}^2\Delta M_{\rm acc})},
\label{eq:delta_s}
\end{align}
in which the second and higher order terms with respect to $\Delta J_{\rm acc}$ and $\Delta M_{\rm acc}$ are ignored in equation~(\ref{eq:delta_s0}) to derive equation~(\ref{eq:delta_s}).
Therefore, the increase or decrease in $r_{\rm sep}$ due to accretion is determined by the sign of the numerator of equation~(\ref{eq:delta_s}).
The binary separation shrinks if the following equation is fulfilled: 
\begin{equation}
    \frac{\Delta J_{\rm acc}}{\Delta M_{\rm acc}} < \frac{3}{2}\frac{J_{\rm b}}{M_{\rm tot}},
\label{eq:delta_s2_org}
\end{equation}
and vice versa. 
With the definition $\Delta j_{\rm acc} \equiv  (\Delta J_{\rm acc}/ \Delta M_{\rm acc}$), equation~(\ref{eq:delta_s2_org}) can be rewritten as  
\begin{equation}
\Delta j_{\rm acc} < \frac{3}{2}\, j_{\rm b}.
\label{eq:delta_s2}
\end{equation}
Thus, roughly speaking, when the specific angular momentum of the accreting matter is smaller or larger than that of the binary system, the binary separation shrinks or widens, respectively. 
However, it is difficult to determine both $\Delta J_{\rm acc}$  and $\Delta M_{\rm acc}$ analytically, because the angular momentum is transported by magnetic braking and the efficiency of  magnetic braking strongly depends on both the envelope mass and the configuration of magnetic field lines, which gradually change with time.  
Thus, in this study, we determine  $\Delta J_{\rm acc}$ and $\Delta M_{\rm acc}$ from  our MHD simulations. 
As described below, we perform this estimate using a sink technique in the simulation.  

\begin{figure*}
\includegraphics[width=0.9\columnwidth]{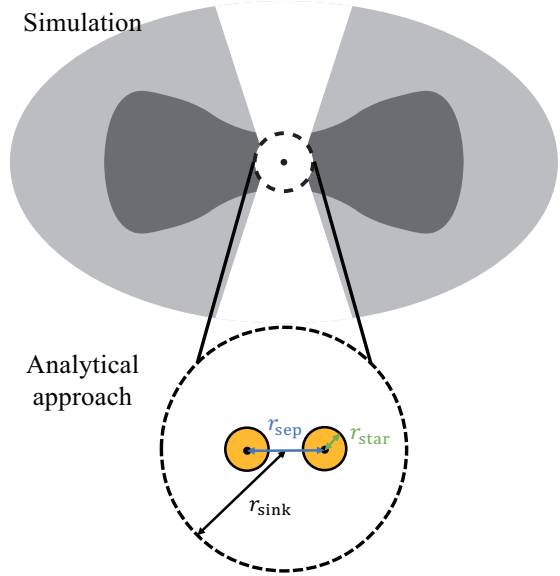}
\caption{
Schematic view of this study. 
We calculate the cloud evolution outside the sink using a 3D MHD simulation. 
After the maximum (or central) density exceeds the sink threshold density, a sink cell is imposed and the accreting mass and angular momentum onto the sink are used to analytically estimate the binary mass and separation. 
We stop the MHD simulation either when the binary separation exceeds twice the sink radius $r_{\rm sep}>2\, r_{\rm sink}$ or when the binary separation is shorter than twice the stellar radius $r_{\rm sep}<2 \, r_{\rm star}$.
}
\label{fig:manga}
\end{figure*}

\subsection{MHD Simulation}
\label{sec:mhdsimulation}
We next explain the MHD simulation. 
Starting from a massive prestellar cloud, we calculate the cloud evolution for $\sim 10^5$\,yr after protostar (or protobinary) formation. 
As the initial state, we adopt a Bonnor--Ebert (BE) density profile, which has a central density of $n_0=10^3\,{\rm cm^{-3}}$ and isothermal temperature of $T_0=20$\,K.
The radius of the initial core is set to twice the critical BE radius, which corresponds to $R_{\rm cl}\simeq3.9\times10^5$\,au.
The density of the BE sphere is enhanced by a factor of $f$ to promote gravitational collapse. 
Thus,  the central density of the initial cloud core is $n_{\rm c,0}=f\times 10^3\, {\rm cm^{-3}}$ \citep[for details, see][]{Matsushita2017}. 
As described in \citet{Matsushita2017}, the mass accretion rate is proportional to $\alpha_0^{-3/2}$, where $\alpha_0$ is the ratio of the thermal to gravitational energy. 
To control the mass accretion rate, we treat $\alpha_0$ as a parameter, with different $f$ values giving different values for $\alpha_0$.
We adopt  $\alpha_0 = 0.2$ and $0.6$, which correspond to $f=3.5$ and $1.2$, respectively. 
Note that the initial cloud mass depends on $\alpha_0$ (or $f$). 

A rigid rotation $\Omega_0$ is taken within the BE sphere.
Three different $\Omega_0$ are adopted to give three different $\beta_0$ (= 0.01, 0.02, and 0.04), the ratio of rotational to gravitational energy. 
We also set a uniform magnetic field $B_0$ over the whole computational domain.
$B_0$ is also treated as a parameter in the range $0$--$5.2\,\mu$G.
The corresponding mass-to-flux ratios normalized by the critical value $(4\pi^2G)^{-1/2}$ are  $\umu_0 =\infty$, $3.0$, $2.0$, and $1.5$, respectively.
Note that the normalized mass-to-flux ratios adopted in this study are typical observed values  \citep[e.g.][]{crutcher10}.
In addition, since it is difficult to form a close binary system with $\mu_0>3$ in our pre-existing simulations, we do not include  the model with $\mu_0>3$.
The model name and parameters are listed in Table~\ref{tab:initial_condition}.
In Cartesian coordinates, the initial direction of the magnetic field is set to be parallel to the $z$-axis, while the initial rotation axis  is inclined from the $z$-axis to the $x$-axis by an angle  $\theta_0$.
It should be noted that the counterclockwise rotation (or positive angular momentum) is adopted around the $x$- and/or $z$-axis.
We also treat $\theta_0$ as a parameter and adopt  $\theta_0=0$, $45$, and $90^\circ$.
Combining  these four parameters ($\alpha_0$, $\beta_0$, $\mu_0$, and $\theta_0$), we prepare 27 models, as listed in Table~\ref{tab:initial_condition}.

To calculate the cloud evolution, we use our nested grid code \citep{machida04,machida05}, which solves the resistive MHD equations (1)--(4) in \citet{Machida2012} with an isothermal equation of state,
\begin{equation}
    P = c^2_s \rho,
\end{equation}
where $c_s$ is the speed of sound.
The diffusion rate for ohmic resistivity is described in \citet{Machida2007} and \citet{Machida2012}.
Grids composed of $(i,j,k)=(64,~64,~64)$ cells are nested and the grid level is described by $l$.
The grid size $L(l)$ and cell width $h(l)$ halve with each increment of the grid level as  $L(l+1) = L(l)/2$ and $h(l+1)=h(l)/2$.
We set six levels of grid, $l=6$, for the initial state.
The initial cloud is immersed in the fifth level of the grid ($l=5$), which has twice the BE radius of $L(5)= 7.85\times10^5\,{\rm au}$ and a cell width of $h(5)=1.23\times10^4\,{\rm au}$.
The coarsest grid has a size of $L(1)=1.26\times10^7\,{\rm au}$ and a cell width of $h(1)= 1.96\times10^5\,{\rm au}$.
After the calculation has been started, a new finer grid is automatically refined to ensure the Truelove condition \citep{Truelove1998}.
In the calculation, the Jeans length is resolved with at least 16 cells.
In the fiducial calculation, we set the maximum grid level to be $l=15$, and the finest grid has $L(15)=767\,{\rm au}$ and $h(15)=12\,{\rm au}$.

\begin{table*}
\centering
\caption{
Model parameters and calculation results. 
Model name, density enhancement factor $f$, cloud mass $M_{\rm cloud}$, ratios of thermal $\alpha_0$ and rotational $\beta_0$ energy to  gravitational energy, mass-to-flux ratio normalized  by the critical value $\mu_0$, magnetic field strength $B_0$, and the angle between rotation axis and magnetic field $\theta_0$ are described.
The ninth column lists the calculation results: models designated wide binary, indicated by `w', close binary, `c', and merger, `m', are described. A `-' entry corresponds to a model for which the binary mass does not reach $20\msun$ within a long calculation period without showing wide binary and merger.
}
    \label{tab:initial_condition}
    \begin{tabular}{ccccccccc}
    \hline
    Model & $f$ & $M_{\rm cloud}$ & $\alpha_0$ & $\beta_0$ &
    $\umu_0$ & $B_0$ & $\theta_0$ & results\\
     & & $[M_{\sun}]$ & & & & ${\rm [\mu G]}$ & $[^{\circ}]$ & \\
    \hline \hline
    T00M20  & \multirow{3}{*}{$3.513$} & \multirow{3}{*}{$262$} &
    \multirow{3}{*}{$0.2$} & \multirow{3}{*}{$0.02$} & \multirow{3}{*}{$2.0$} &
    \multirow{3}{*}{$3.9$} & 0 & -\\
    T45M20  & & & & & & & 45 & w\\
    T90M20  & & & & & & & 90 & c\\
    \hline
    T00M30  & \multirow{3}{*}{$3.513$} & \multirow{3}{*}{$262$} &
    \multirow{3}{*}{$0.2$} & \multirow{3}{*}{$0.02$} & \multirow{3}{*}{$3.0$} &
    \multirow{3}{*}{$2.6$} & 0 & w\\
    T45M30  & & & & & & & 45 & w\\
    T90M30  & & & & & & & 90 & c\\
    \hline
    T00M15  & \multirow{3}{*}{$3.513$} & \multirow{3}{*}{$262$} &
    \multirow{3}{*}{$0.2$} & \multirow{3}{*}{$0.02$} & \multirow{3}{*}{$1.5$} &
    \multirow{3}{*}{$5.2$} & 0 & m\\
    T45M15  & & & & & & & 45 & c\\
    T90M15  & & & & & & & 90 & c\\
    \hline
    T00M20A6  & \multirow{3}{*}{$1.171$} & \multirow{3}{*}{$87$} &
    \multirow{3}{*}{$0.6$} & \multirow{3}{*}{$0.02$} & \multirow{3}{*}{$2.0$} &
    \multirow{3}{*}{$1.3$} & 0 & w\\
    T45M20A6  & & & & & & & 45 & w\\
    T90M20A6  & & & & & & & 90 & -\\
    \hline
    T00M30A6  & \multirow{3}{*}{$1.171$} & \multirow{3}{*}{$87$} &
    \multirow{3}{*}{$0.6$} & \multirow{3}{*}{$0.02$} & \multirow{3}{*}{$3.0$} &
    \multirow{3}{*}{$0.86$} & 0 & w\\
    T45M30A6  & & & & & & & 45 & w\\
    T90M30A6  & & & & & & & 90 & -\\
    \hline
    T00M15A6  & \multirow{3}{*}{$1.171$} & \multirow{3}{*}{$87$} &
    \multirow{3}{*}{$0.6$} & \multirow{3}{*}{$0.02$} & \multirow{3}{*}{$1.5$} &
    \multirow{3}{*}{$1.7$} & 0 & m\\
    T45M15A6  & & & & & & & 45 & -\\
    T90M15A6  & & & & & & & 90 & -\\
    \hline
    T00M20B1  & \multirow{3}{*}{$3.513$} & \multirow{3}{*}{$262$} &
    \multirow{3}{*}{$0.2$} & \multirow{3}{*}{$0.01$} & \multirow{3}{*}{$2.0$} &
    \multirow{3}{*}{$3.9$} & 0 & -\\
    T45M20B1  & & & & & & & 45 & c\\
    T90M20B1  & & & & & & & 90 & c\\
    \hline
    T00M20B4  & \multirow{3}{*}{$3.513$} & \multirow{3}{*}{$262$} &
    \multirow{3}{*}{$0.2$} & \multirow{3}{*}{$0.04$} & \multirow{3}{*}{$2.0$} &
    \multirow{3}{*}{$3.9$} & 0 & w\\
    T45M20B4  & & & & & & & 45 & w\\
    T90M20B4  & & & & & & & 90 & w\\
	\hline
	T00MinfB1 & \multirow{3}{*}{$3.513$} & \multirow{3}{*}{$262$} &
	\multirow{3}{*}{$0.2$} & $0.01$ & \multirow{3}{*}{$\infty$} &
	\multirow{3}{*}{$0$} & - & w\\
	T00Minf & & & & $0.02$ & & & - & w\\
	T00MinfB4 & & & & $0.04$ & & & - & w\\
	\hline
    \end{tabular}
\end{table*}

\subsection{Mass and Angular Momentum Introduced into Sink}
\label{sec:sink}
As described in \S\ref{sec:separation},  we can investigate the evolution of  the binary separation when the accretion rate of the mass and angular momentum are known. 
To realize a long-time evolution, we adopt a sink cell  \citep[for details, see][]{MachidaHosokawa2013}.
We set the sink at the center of the maximum level of the grid ($l=l_{\rm max}$).
We describe the sink threshold density as $n_{\rm sink}$ and the sink radius as $r_{\rm sink}$. 
Note that the sink begins to operate after the central density exceeds $n_{\rm sink}$.  
We treat the mass within the sink as a point mass in the simulation.
Using the number density $n_i$, distance from the center $\bm{r}_i$, velocity $\bm{v}_i$, and volume ${\rm dV}_i$ at each cell within $r_{\rm sink}$,  we estimate the accreting mass $M_{\rm sink}$ and  angular momentum $\bm{J}_{\rm sink}$ at every step of the simulation as
\begin{align}
\Delta M_{\rm acc} &= \sum_{r_i < r_{\rm sink}} {\rm C_{acc}} ~\bar{\umu} m_{\rm p} (n_i - n_{\rm sink}) ~{\rm dV}_i, \\
\Delta \bm{J}_{\rm acc} &= \sum_{r_i < r_{\rm sink}} {\rm C_{acc}} ~\bar{\umu} m_{\rm p} (n_i - n_{\rm sink})
~\bm{r}_i \times \bm{v}_i ~{\rm dV}_i,
\end{align}
where $\bar{\umu} (=2.4) $ and $m_{\rm p}$ are the mean molecular weight and proton mass, respectively. 
The accretion factor ${\rm C_{acc}}$, which smoothly changes the structure within the sink, is set to ${\rm C_{acc}} = 0.03$. 
We confirmed that the results are not significantly affected by the accretion factor ${\rm C_{acc}}$. 
For the fiducial models, we set $n_{\rm sink}=10^9\cm$, and the sink radius is defined as half of the Jeans length $\lambda_{\rm J}$ as
\begin{equation}
	r_{\rm sink} = \frac{1}{2} \lambda_{\rm J} \sim 66\,{\rm au}.
	\label{eq:r_sink}
\end{equation}
Thus, $n_{\rm sink}$ is related to $r_{\rm sink}$ through $\lambda_{\rm J}$.
We change $n_{\rm sink}$ and the corresponding $r_{\rm sink}$ in some calculations to investigate the dependence of our results on the spatial resolution (for details, see \S\ref{sec:resoluton}).

To determine the binary separation, the mass and angular momentum of accreting matter are updated as  
\begin{align}
M_{\rm tot, new} &= M_{\rm tot, old} + \Delta M_{\rm acc},  \\
\bm{J}_{\rm b, new} &= \bm{J}_{\rm b, old} + \Delta {\bm J_{\rm acc}},
\label{eq:sink}
\end{align}
where $M_{\rm tot, old}$ and $\bm{J}_{\rm b, old}$ are the mass and angular momentum of the binary system at the previous step, and $M_{\rm tot, new}$ and $\bm{J}_{\rm b, new}$ are those at the present step, respectively. 
Using these equations, we determine the binary separation with equation~(\ref{eq:separation3}).
Note that we use the absolute value of $J_{\rm b} = \vert \vect{J}_{\rm b, new} \vert  $ when estimating the binary separation with equation~(\ref{eq:separation2}).

\subsection{Stellar Evolution and Termination Condition of Simulation}
\label{sec:stellarevolution}
In this subsection, we describe the limitations of estimating the binary separation and the termination condition in our simulations. 
As described above, an equal mass binary system is assumed within the sink in our analysis, and the binary separation is estimated at every time step with the mass and angular momentum falling into the sink. 
Thus, when the binary separation exceeds the diameter of the sink (i.e. $r_{\rm sep}> 2 r_{\rm sink}$), we cannot adequately estimate the separation. 
In other words, if the binary system exists outside the sink, we cannot estimate the mass and angular momentum introduced into the binary system because the mass accretion is assumed to occur only onto the sink. 
We estimate the binary separation according to the procedure described in \S\ref{sec:sink} only when $r_{\rm sep}$ is shorter than the sink diameter $2\, r_{\rm sink}$ (i.e. $r_{\rm sep}<2\,r_{\rm sink}$).
Thus, twice the sink radius $2\, r_{\rm sink}$ is the upper limit of the binary separation that can be estimated in this study. 
We stop the simulation when $r_{\rm sep}> 2\, r_{\rm sink}$ is reached, and we call a model exceeding the upper limit a `wide binary.'

We also need to impose a lower limit of the binary separation. 
Each protostar in a binary system has a physical size or radius.
We calculate the radius of each protostar $r_{\rm star}$ by numerically solving the stellar interior structure. 
As already described, we assume that the two protostars accrete the gas equally distributed among them. We construct an accretion history $\Delta M_{\rm acc}/2$ as a function of $M_{\rm tot}/2$ taken from MHD simulation data.
We employ the same numerical code as in \citet{MachidaHosokawa2013}, assuming spherical accretion onto the protostar \citep[][]{Hosokawa2009}. While we use a different stellar evolution code in \cite{Matsushita2018} and \cite{Machida2020}, an advantage of the current method is that the evolution of the stellar radius only depends on the accretion history, not on other control parameters \citep[see Section~3 in][]{Matsushita2018}. The details in modeling the protostellar evolution do not affect our conclusions anyway \citep[e.g.][]{Hosokawa2010}. 
When the binary separation is smaller than  twice the stellar radius (i.e. $r_{\rm sep}<2\, r_{\rm star}$), we determine that merger of the two protostars has occurred,
and  the calculation is stopped.
We call a model exhibiting merger a `merger.'

Hence, as long as the binary separation is in the range $2\, r_{\rm star} < r_{\rm sep} <2\, r_{\rm sink}$, we continue the calculation until $M_{\rm tot}$ reaches $20\msun$ (i.e. the mass of each protostar is  $10\msun$).
We call such a model a `close binary,' in which the binary system maintains the separation ranging from $2\, r_{\rm star}$ to $ 2\, r_{\rm sink}$ until the end of the simulation ($M_{\rm tot}=20\msun$).
It takes several months to perform the calculations for a single model.
We stop the calculation if the total mass does not reach  $M_{\rm tot}=20\msun$ within about four months of wall-clock time, roughly corresponding to $1.2\times10^4$ CPU hours.
Thus, in some models,  the total mass is less than $M_{\rm tot}<20\msun$ even when the binary separation is in the range $2\, r_{\rm star} < r_{\rm sep} <2\, r_{\rm sink}$ during the calculation.  
A schematic view of our settings and the limitation of our simulation is presented in Figure~\ref{fig:manga}, in which $r_{\rm sink}$, $r_{\rm sep}$, and the stellar radius $r_{\rm star}$ are described.

\subsection{Summary of Our Procedure}

To determine the binary parameters such as total protostellar mass and separation, we need information about the accretion rates of the mass and angular momentum, which can be  determined from the 3D MHD simulations. 
Thus, we can directly calculate the configuration and strength of the magnetic field on a large scale with 3D simulations. 
In other words, in the simulations, we estimate the mass and angular momentum falling into the binary system using the sink technique,
and then use this information to estimate the mass and separation of the binary system. 

A major difference between our study and \citet{Lund2018} is the treatment of the magnetic field. 
We determine the accretion history of the mass and angular momentum from our MHD simulation, while  \citet{Lund2018} analytically estimated them. 
Thus, our treatment of  the magnetic field is more precise than the previous work.  
On the other hand,  the spatial scale in our study is limited to the sink radius (see \S\ref{sec:stellarevolution} and \ref{sec:resolution}), while there is no limitation of the spatial scale in the analytical study of \citet{Lund2018}.

\section{Results}
\label{sec:results}
\subsection{Fiducial Model} \label{sec:typicalmodel}

\begin{figure*}
\includegraphics[width=0.9\columnwidth]{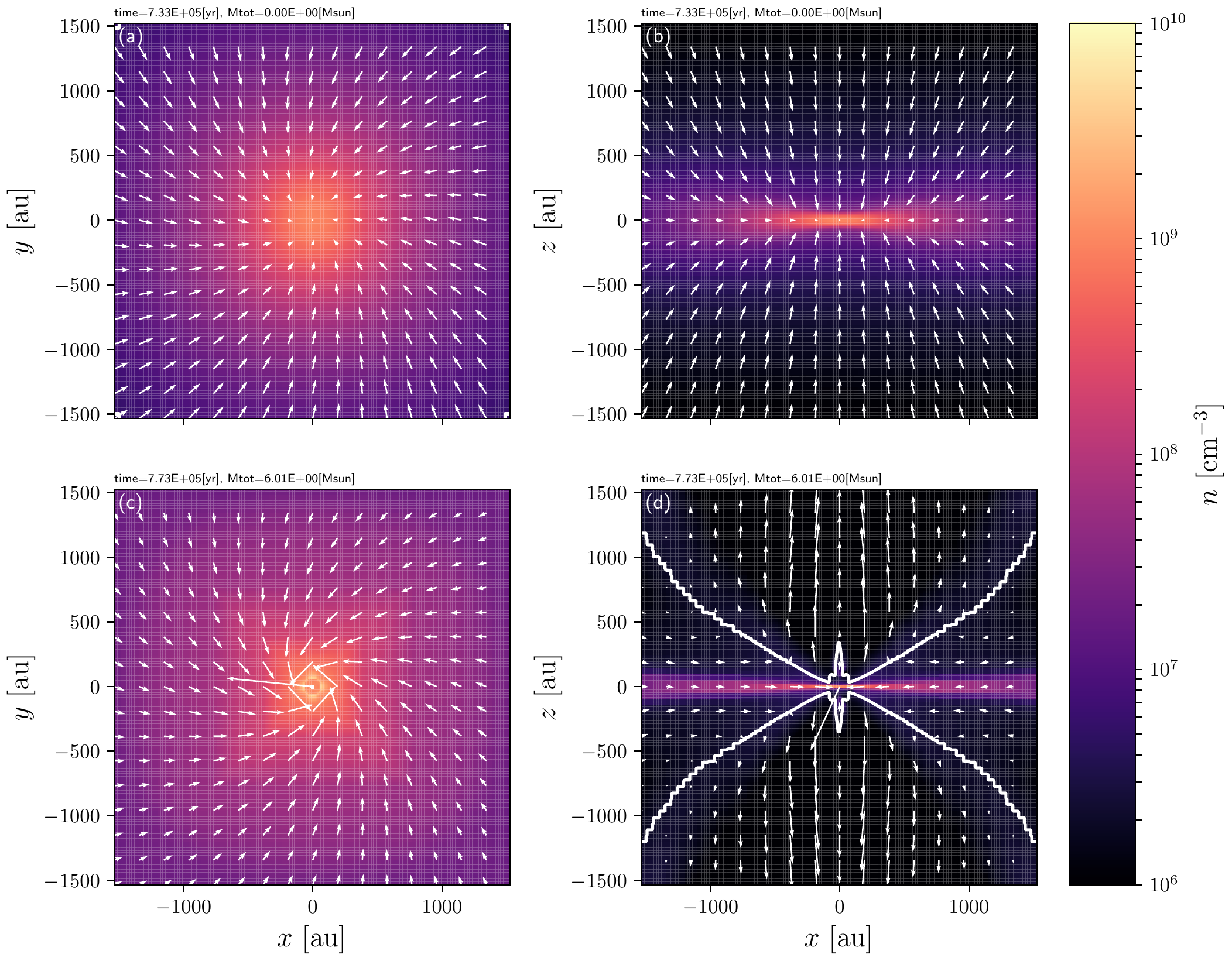}
\caption{
Density (color) and velocity (arrows) distributions on the $z=0$ (left panels) and $y=0$ (right panels) planes just before (top panels) and $4\times10^4$\,yr after (bottom panels) binary formation for model T00M20.
The binary mass and elapsed time from the beginning of the cloud collapse are given in each panel.
The white contour in panel (d) corresponds to the boundary between the infalling $v_r < 0$ and outflowing $v_r>0$ gas.
}
\label{fig:colormap}
\end{figure*}

Firstly, we describe the results for model T00M20, which has the parameters $\alpha_0=0.2$, $\beta_0=0.02$, $\mu_0=2.0$, and $\theta_0=0^\circ$. 
We call this the fiducial model. 
Other models are shown in \S\ref{sec:parameter}.

\begin{figure*}
\includegraphics[width=0.9\columnwidth]{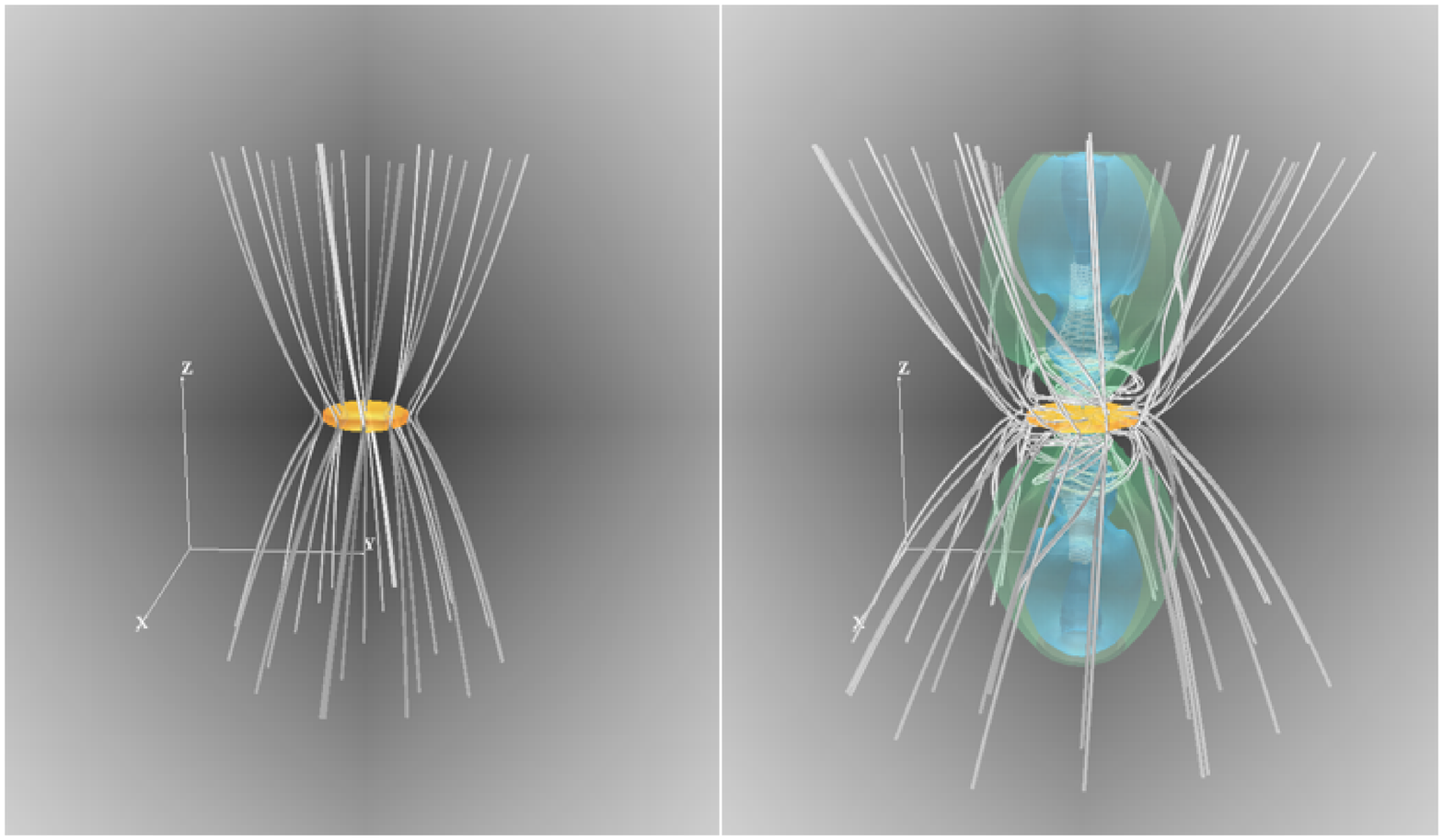}
\caption{
Three dimensional views for model T00M20 at the same epochs as in Fig.~\ref{fig:colormap}. 
The white lines correspond to magnetic field lines.
In each panel, the pseudo disk is represented by the yellow surface, which correspond to the iso-density surface of $5.0\times10^6$\,cm$^{-3}$.
In the right panel, the outflow is represented by the green and blue surfaces, which correspond to the iso-velocity surfaces of $v_r=0.13$ and $2.6$\,km\,s$^{-1}$, respectively.  
Within the green surface,  the gas is outflowing from the center.
The box size is $\sim 10^4$\,au in both panels.
}
\label{fig:ap2t00_bfield}
\end{figure*}

\subsubsection{Density Distribution and Configuration of Magnetic Field}
\label{density_map}
Figure~\ref{fig:colormap} shows the density and velocity distributions before and after binary formation for the fiducial model (T00M20).
In the edge-on view (right panels of Fig.~\ref{fig:colormap}), a thin disk-like structure can be seen.
In Figure~\ref{fig:colormap}{\it b},  the gas is flowing into the sink, while there is no outflowing component.  
On the other hand, Figure~\ref{fig:colormap}{\it d} shows an outflow driven from the disk-like structure. 
Note that the disk-like structure corresponds to the pseudo disk that is mainly supported against gravity by the Lorentz force, and thus the disk is not supported by the rotation, as  seen in Figure~\ref{fig:colormap}{\it c}. 
In this study, we did not resolve a rotationally supported disk, because the calculation was stopped before the formation of a rotationally supported disk as described in \S\ref{sec:stellarevolution}. 
Nevertheless, the outflow appears because the magnetic field lines are twisted by the (sub-Keplerian) rotation of the pseudo disk and the magnetic pressure gradient force drives the outflow (see Figs.~\ref{fig:colormap}{\it c} and {\it d}).
We show the outflow evolution and the relation between the outflow and binary separation in \S\ref{sec:outflow}.

The configuration of the magnetic field lines and the structure of the pseudo disk and outflow for the fiducial model (T00M20) are plotted in Figure~\ref{fig:ap2t00_bfield}. 
In the left panel of Figure~\ref{fig:ap2t00_bfield} (before the sink creation), although an hourglass configuration of the magnetic field lines is confirmed, the magnetic field lines are not strongly convergent toward the center.    
Meanwhile, the magnetic field lines are highly convergent and strongly twisted a long time after sink creation (Fig.~\ref{fig:ap2t00_bfield} right panel). 
In other words, the magnetic field lines are nearly parallel just before binary formation (Fig.~\ref{fig:ap2t00_bfield} left), while they show a fan-shaped configuration long after binary formation. 
The efficiency of angular momentum transport  due to  magnetic braking strongly depends on 
the configuration of the magnetic field lines \citep{Hirano2020}. 
Since the configuration of magnetic field lines changes over time, the binary separation changes accordingly (see \S\ref{sec:parameter_theta}).

\begin{figure}
\begin{center}
\includegraphics[width=0.9\columnwidth]{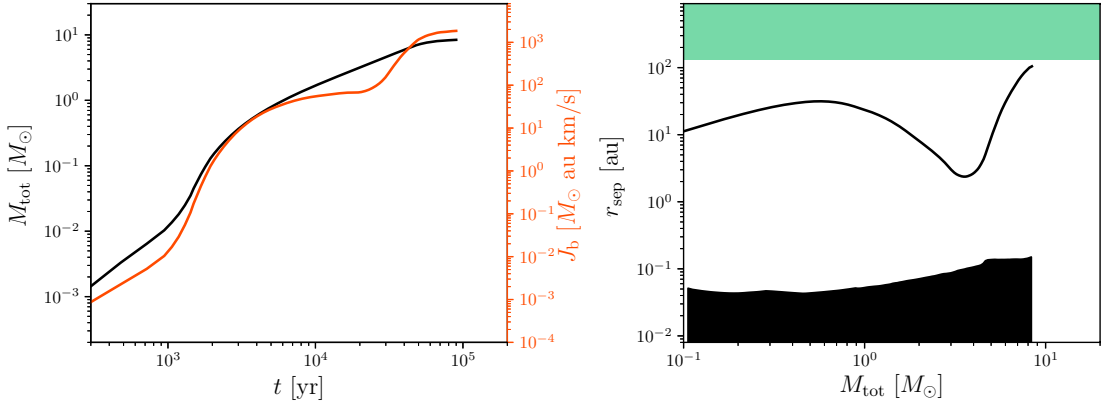}
\caption{
Left:
Total mass $M_{\rm tot}$ (black; left axis)  and angular momentum $J_{\rm b}$ (red; right axis) falling onto the sink against the elapsed time from the formation of the binary $t$ for model T00M20.
Right:
Binary separation against the total stellar mass for model T00M20.
The boundary between the green and white areas corresponds to twice the sink radius (or the upper limit of binary separation applicable for this study). 
The boundary between the white and black areas corresponds to twice the stellar radius (or the lower limit of binary separation) for model T00M20.
}
\label{fig:main-results}
\end{center}
\end{figure}

\subsubsection{Binary Mass, Angular Momentum, and Separation}

The left panel of Figure~\ref{fig:main-results} shows the time evolution of the total mass $M_{\rm tot}$ and angular momentum $J_{\rm b}$ falling onto the sink for the fiducial model (T00M20).
The right panel of Figure~\ref{fig:main-results} shows the binary separation estimated using equations~(\ref{eq:separation2})--(\ref{eq:delta_s}).
For this model, we stopped the calculation just before the binary separation reaches the upper limit (twice the sink radius), because the calculation timestep becomes very short due to the emergence of the high-speed outflow (Fig.~\ref{fig:ap2t00_bfield}).

Figure~\ref{fig:main-results} left panel indicates that both the total mass  and angular momentum continue to increase until the end of the simulation. 
The wavy curve in $J_{\rm b}$ is caused by outflow emergence, which is explained in \S\ref{sec:outflow}.
As shown in the right panel of Figure~\ref{fig:main-results}, the binary separation slowly increases for  $M_{\rm tot} \lesssim  0.5\msun$,  decreases for  $0.5\msun \lesssim M_{\rm tot}\lesssim 4\msun$, and  
then increases again, reaching $\sim 100$\,au at the end of the simulation. 
The curious behavior of the binary separation for model T00M20 is explained further in \S\ref{sec:outflow}.

\begin{figure*}
\includegraphics[width=0.9\columnwidth]{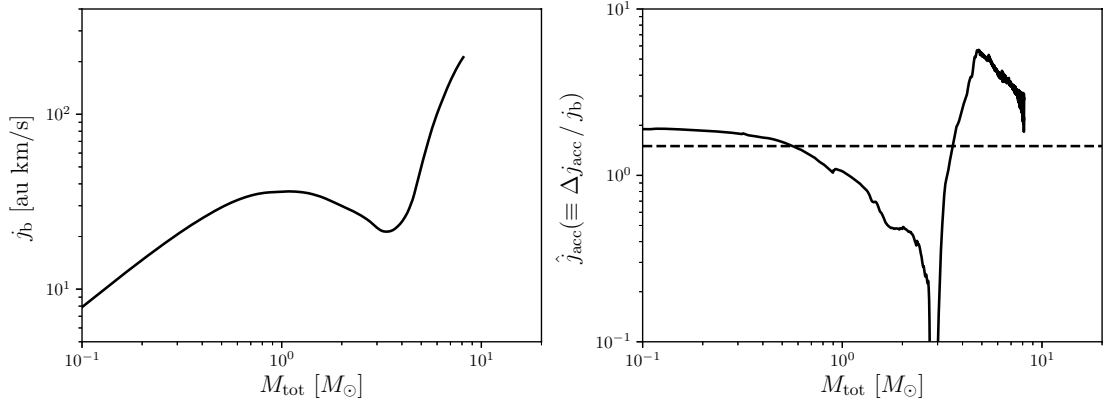}
\caption{
Specific angular momentum $j_{\rm b} \equiv J_{\rm b}/M_{\rm tot}$ (left panel) and ratio  $\Delta j_{\rm acc}/j_{\rm b}$ (right panel) against the total mass for model T0020.
The broken line in the right panel corresponds to $\hat{j}_{\rm acc}=3/2$.
}
\label{fig:main_results_multi}
\end{figure*}

\subsubsection{Critical Specific Angular Momentum}
The specific angular momentum of the binary system ($j_{\rm b}=J_{\rm b}/M_{\rm tot}$) for the fiducial model (T00M20) is plotted in the left panel of Figure~\ref{fig:main_results_multi}, and shows a similar tendency to  the total angular momentum (or the angular momentum of the binary system) $J_{\rm b}$ (Fig.~\ref{fig:main-results} left panel).
The ratio of the specific angular momentum of the accreting matter to the specific angular momentum of the binary system ($\Delta j_{\rm acc}/j_{\rm b}$) for the fiducial model is also plotted in the right panel of Figure~\ref{fig:main_results_multi}.
In the following, we describe the ratio  $\Delta j_{\rm acc}/j_{\rm b}$ as $\hat{j}_{\rm acc}$ and call it the accreting angular momentum.
As described in \S\ref{sec:separation}, we can speculate the time evolution of the binary separation with $\hat{j}_{\rm acc}$ (see eq.~[\ref{eq:delta_s2}]). 
The binary separation decreases with  $ \hat{j}_{\rm acc}<3/2 $ (hereafter we call  $ \hat{j}_{\rm acc}=3/2$  the critical angular momentum, $\hat{j}_{\rm cri}$).  
Thus, the binary separation shrinks when the angular momentum of the infalling matter is sufficiently removed and $ \hat{j}_{\rm acc} < \hat{j}_{\rm cri}$. 
On the other hand, the condition $ \hat{j}_{\rm acc}> \hat{j}_{\rm cri}$ should be realized when the angular momentum transport in the infalling matter is not efficient,  because the specific angular momentum later falling onto the sink is large, as described above.
Without an efficient mechanism for angular momentum transport, the binary separation continues to widen with time. 

A comparison of Figure~\ref{fig:main-results} right panel and  Figure~\ref{fig:main_results_multi} right panel shows that the binary separation $r_{\rm sep}$ is well correlated with the accreting angular momentum $\hat{j}_{\rm acc}$.  
Actually, the binary separation widens when the accreting angular momentum $\hat{j}_{\rm acc}$ is larger than the critical angular momentum $\hat{j}_{\rm cri}$ and vice versa. 
Thus, we can expect the binary separation to evolve with the accreting angular momentum.

\subsubsection{Effect of Outflow on Binary Separation}
\label{sec:outflow}
We applied a large sink radius ($r_{\rm sink}=66$\,au) in this study. 
We set the sink radius so as to not resolve both the rotationally supported disk and binary orbital motion, in order to investigate the angular momentum falling into the central region.   
Although no rotationally supported disk appears in this study, outflow, which is usually driven by a rotationally supported disk, appears, as shown in Figures~\ref{fig:colormap} and \ref{fig:ap2t00_bfield}. 
The outflow driving is realistic, because the sub-Keplerian motion of the pseudo disk can twist the magnetic field lines and produce  a magnetic pressure gradient to drive the outflow, as described in \S\ref{density_map}.

As shown in Figure~\ref{fig:colormap}, a noticeable outflow can be seen in the aligned case ($\theta_0=0^\circ$). 
Although outflow appears in the misaligned cases ($\theta_0=45$ and 90$^\circ$, see \S\ref{sec:oscillation}), it is not very clear and is not strong,  which is consistent with previous studies \citep{lewis15, lewis17, Hirano2020}.
In this subsection, we describe the relation between the outflow driving and the accreting angular momentum, focusing on the aligned case (or the fiducial model T00M20) because a strong outflow appears for $\theta_0=0^\circ$.

\begin{figure*}
\centering
\includegraphics[width=0.9\columnwidth]{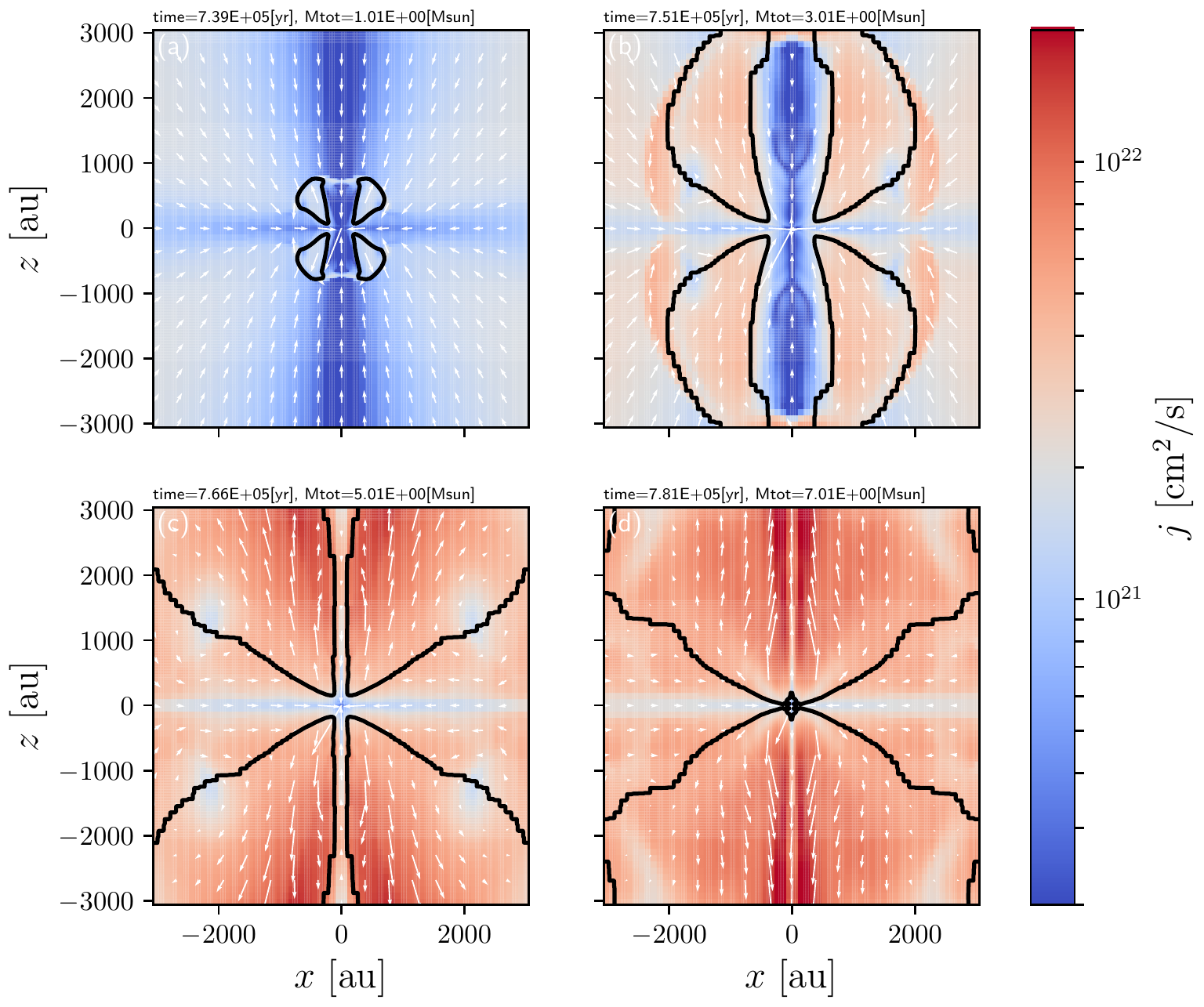}
\caption{
Distributions  of the specific angular momentum (color) and velocity (arrows) on the $y=0$ plane at four different epochs when the sink mass reaches $M_{\rm tot}=1$, $3$, $5$, and $7 M_\odot$ for model T00M20.  
The boundary between the outflowing and infalling region, which corresponds to $v_r=0$, is indicated by the black contour. 
The elapsed time from the beginning of the cloud collapse and total stellar mass are given in each panel.
}
\label{fig:j_profile}
\end{figure*}

Figure~\ref{fig:j_profile} shows the distribution of the specific angular momentum in both regions outside and inside the outflow. 
The figure indicates that the specific angular momentum inside the outflow is larger than that outside the outflow. 
Thus, a significant amount of angular momentum is transferred with the outflow, in addition to magnetic braking.  
Moreover, from the figure, we can see that the outflow has a wide opening angle in the late phase during which  the gas accretion is significantly  obstructed by the outflow. 

Without the outflow, the gas distributed along the $z$-axis would preferentially accrete onto the central region, because the (specific) angular momentum of such gas is small. 
On the other hand, as seen in the bottom panels of Figure~\ref{fig:j_profile}, the outflow interrupts the accretion of the gas distributed just above and below the central region (i.e. sink or proto-binary region).
The gas distributed near the rotation axis (or $z$-axis), which has a small amount of angular momentum, is swept away by the outflow and cannot accrete onto the central region. 
As a result, only the gas with large angular momentum distributed in the region outside the outflow can fall onto the central region in the late main accretion phase. 
It should be noted that, in the bottom panels of Figures~\ref{fig:j_profile}, the angular momentum inside the outflow is larger than that outside the outflow. 
Especially, the angular momentum along the $z$-axis is the largest inside the outflow. 
However, if no outflow appears, the gas with a small amount of angular momentum would fall onto the central region along the $z$-axis, as seen in Figure~\ref{fig:j_profile}{\it a}. 
Hence, the outflow interrupts the accretion of the gas with a small amount of angular momentum.

As seen in Figure~\ref{fig:main_results_multi}, for the fiducial model (T00M20), the specific angular momentum and the separation begin to decrease around $M_{\rm tot}\sim 1\msun$, and then increase again for $M_{\rm tot}\gtrsim 3\msun$.
The decrease of the angular momentum around $M_{\rm tot}\sim1\msun$ can be attributed to a change in the configuration of the magnetic field, as described in \S\ref{density_map} and \S\ref{sec:parameter_theta}.
Since  magnetic braking is more efficient in the fan-shaped configuration than in the parallel configuration, the angular momentum is largely removed from the center for the fan-shaped configuration, which is realized in the late accretion phase \citep{tsukamoto18,Hirano2020}.
Meanwhile,  only gas with large angular momentum is introduced into the central region after the outflow is sufficiently matured, as described above.
As a result,  the angular momentum of the binary system increases. 

Thus, when the angular momentum removed by the outflow and magnetic braking is smaller than that  of the accreting matter, the binary separation will gradually widen with a matured outflow. 
In summary, the increase and decrease of the angular momentum and binary separation is related to both the configuration of the magnetic field and outflow growth.

\begin{figure*}
\includegraphics[width=0.9\columnwidth]{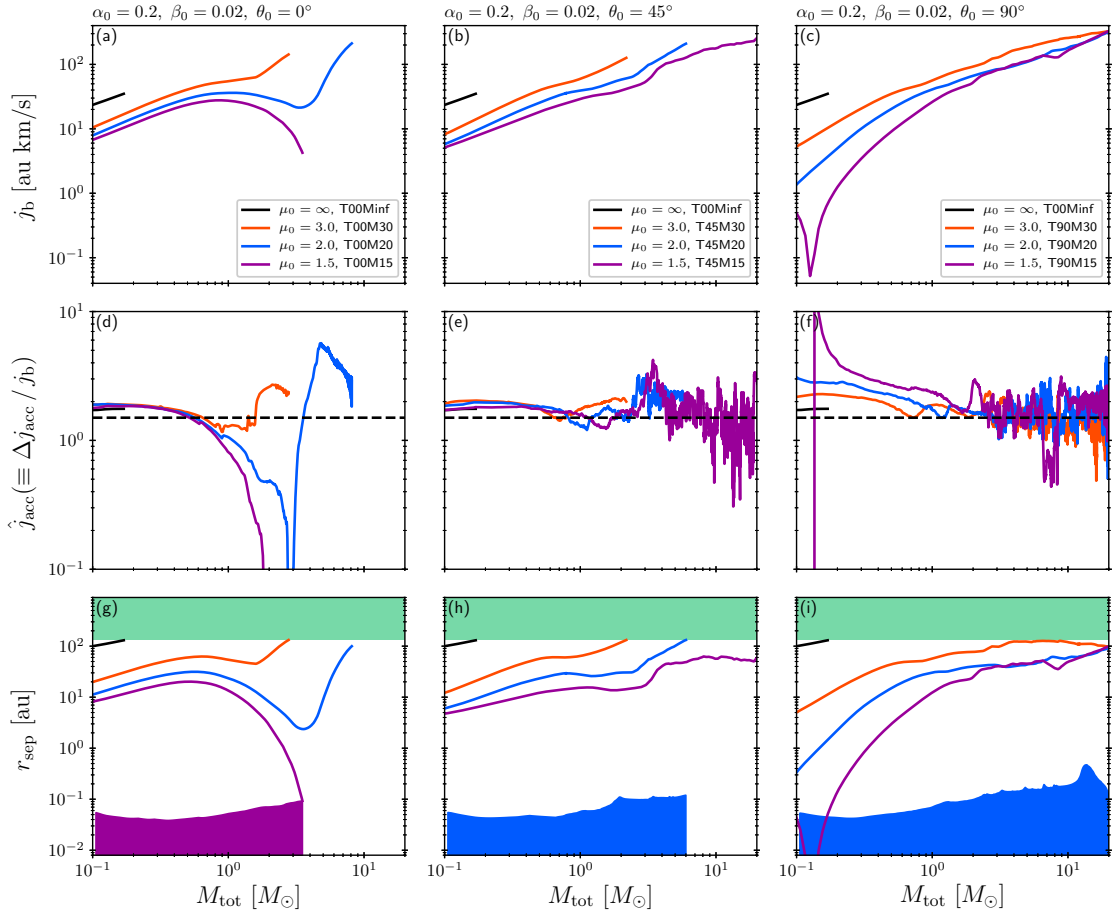}
\caption{
Binary specific angular momentum $j_{\rm b}$ (top panels), accreting angular momentum $\hat{j}_{\rm acc}$ (middle panels), and binary separation $r_{\rm sep}$ (bottom panels) against total stellar mass $M_{\rm tot}$.  
Models with different $\mu_0$ but with the same  $\theta_0$  are plotted in each panel. 
$\alpha_0$, $\beta_0$, and $\theta_0$ are given in each top panel.
The broken line in the middle panels indicates $\hat{j}_{\rm cri}$. 
In the bottom panels, the boundary between the lower colored and white areas corresponds to the lower limit of binary separation (twice the stellar radius), in which the evolution of the stellar radius is calculated with the accretion history for the model having the same color.
Since there is no significant difference in the evolution of the stellar radii in each panel, only the radii for the models with $\mu_0=1.5$ ({\it g}) and $\mu_0=2.0$ ({\it h} and {\it i}) are plotted.
The green area in the bottom panels is the region outside the sink.  
}
\label{fig:compare_mu}
\end{figure*}

\subsection{Parameter Dependence}
\label{sec:parameter}
In this section, we describe the results for all  models listed in Table~\ref{tab:initial_condition}.

\subsubsection{Dependence on Initial Magnetic Field Strength}
\label{sec:paramag}
We investigate the dependence of  the binary separation on the magnetic field strength  in this subsection.
Figure~\ref{fig:compare_mu} shows the evolution of $j_{\rm b}$ (top panels), $\hat{j}_{\rm acc}$ (middle panels), and  $r_{\rm sep}$  (bottom panels) with different magnetic field strengths $B_0$ or $\mu_0$
\footnote{
We apply a special treatment to model T90M15.
For this model, the binary separation is smaller than twice the stellar radius for $M_{\rm tot} \lesssim 0.15\msun$ (Fig.~\ref{fig:compare_mu}{\it i}) during which the angular momentum is also considerably small (Fig.~\ref{fig:compare_mu}{\it c}), and thus it is judged that a merger occurs at a very early epoch  ($M_{\rm tot}\simeq0.15\msun$)  in our definition. 
However, we can assume that the binary formation (or fragmentation) occurs after $M_{\rm tot}>0.15\msun$ because  the angular momentum and binary separation continue to increase in the following stage (Figs.~\ref{fig:compare_mu}{\it c} and \ref{fig:compare_mu}{\it i}). 
Thus, for model T90M15, we only consider the binary evolution for $M_{\rm tot}>0.2\msun$.
}. 

There is a clear trend for small $j_{\rm b}$  and $r_{\rm sep}$ in the models with a relatively strong magnetic field ($\mu_0=1.5$) compared to models with a relatively weak magnetic field ($\mu_0=3.0$), which is consistent with \citet{Lund2018}.
In addition, $j_{\rm b}$  and $r_{\rm sep}$ in the unmagnetized model are much larger than in the magnetized models.
Thus, the effect of  magnetic field on the angular momentum transfer and binary separation is clear.

For the models with $\theta_0=0^\circ$ (Fig.~\ref{fig:compare_mu} left panels), the difference in the separation between the models is  small in the early phase ($M_{\rm tot} \lesssim 1\msun$), while it  becomes significant in the late phase ($M_{\rm tot} \gtrsim 1\msun$), as seen in Figure~\ref{fig:compare_mu}{\it g}.
Especially, for the model with $\mu_0=1.5$, $r_{\rm sep}$ becomes smaller than twice the stellar radius (or below the lower limit). 
As described in \S\ref{sec:parameter_theta}, a rapid decrease in the separation can be attributed to a change of the configuration of the magnetic field (see also \citealt{Hirano2020}). 
The outflow also contributes to the evolution of the binary separation and angular momentum, which is shown  in \S\ref{sec:outflow}.
Meanwhile, for the models with $\theta_0=90^\circ$ (Fig.~\ref{fig:compare_mu}{\it i}), the difference in $r_{\rm sep}$ is large in the early phase ($M_{\rm tot} \lesssim 1\msun$). 
However, the difference is not significant at the end of the simulation ($M_{\rm tot}=20\msun$), which indicates that the efficiency of  magnetic braking does not strongly depend on the initial magnetic field strength when the magnetic field vectors are not aligned with the rotation axis (for details, see \S\ref{sec:parameter_theta}).

The evolution of $r_{\rm sep}$ for the models with $\theta_0=45^\circ$ (Fig.~\ref{fig:compare_mu}{\it h}) is similar to that for the perpendicular models $\theta_0=90^\circ$ (Fig.~\ref{fig:compare_mu}{\it i}), while there exist quantitative differences in $j_{\rm b}$, $\hat{j}_{\rm acc}$ and $r_{\rm sep}$ between the models.
In the models with $\theta_0=45$ and $90^\circ$,  $j_{\rm b}$  (Figs.~\ref{fig:compare_mu}{\it b} and {\it c}) and the accreting angular momentum  $\hat{j}_{\rm acc}$ (Figs.~\ref{fig:compare_mu}{\it e} and {\it f}) roughly trace a similar track, although they are not the same in the very early phase.
In the late phase, $j_{\rm b}$ continues to increase, while $\hat{j}_{\rm acc}$ oscillates around $\hat{j}_{\rm cri}$ after it decreases to reach $\hat{j}_{\rm cri}$. 
Note that, for model T45M30, since the binary separation reaches the upper limit at an early epoch, the oscillation in $\hat{j}_{\rm acc}$ cannot be seen. 
We discuss the oscillation of $\hat{j}_{\rm acc}$ in \S\ref{sec:oscillation}.
On the other hand, $\hat{j}_{\rm acc}$ behaves differently for the aligned models $\theta_0=0^\circ$ (Fig.~\ref{fig:compare_mu}{\it d}), in which $\hat{j}_{\rm acc}$ for the models with $\mu_0=1.5$ and 2 rapidly drops at $M_{\rm tot}\sim0.5\msun$ and then increases to $\hat{j}_{\rm acc} \sim \hat{j}_{\rm cri}$ for $M_{\rm tot} \gtrsim 3$--$5\msun$.
Reflecting the complicated  behavior of $\hat{j}_{\rm acc}$, the evolution of $r_{\rm sep}$ is not simple  for the aligned models (Fig.~\ref{fig:compare_mu}{\it g}).  
Thus, the aligned models may be a special case for the angular momentum transport and evolution of the binary separation. 

\begin{figure*}
\includegraphics[width=0.9\columnwidth]{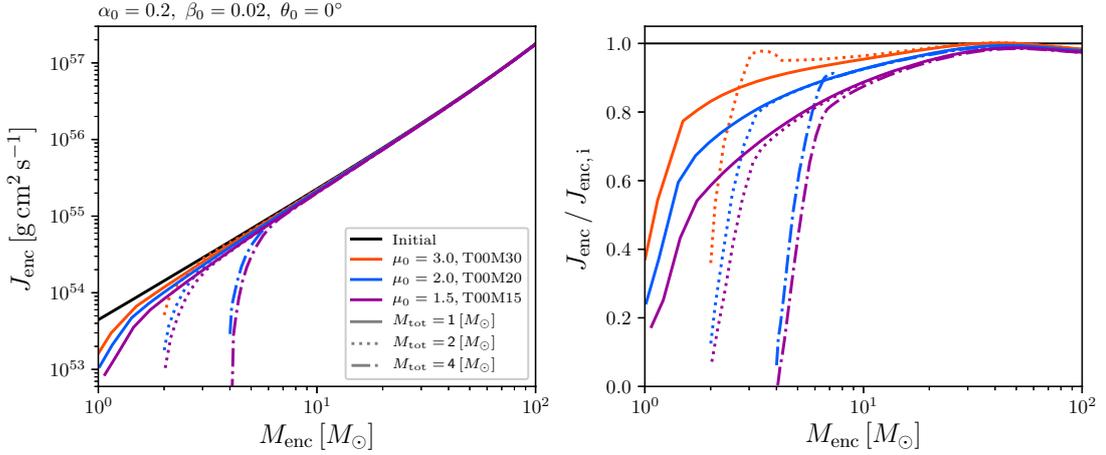}
\caption{
Distribution of  enclosed angular momentum (left) and enclosed angular momentum normalized by the initial value (right) against the enclosed mass for models with different $\mu_0$ at the epochs when the binary mass reaches $M_{\rm tot}=1$, $2$, and $4\,M_\odot$.
The black solid line in both panels is the initial distribution. 
The parameters $\alpha_0$, $\beta_0$, and $\theta_0$ are given in the upper part of left panel.
}
\label{fig:compare_mu_profile}
\end{figure*}

Finally, we confirm the distribution of the angular momentum after binary  formation (or sink creation). 
Figure~\ref{fig:compare_mu_profile} left panel shows the distribution of the enclosed angular momentum  at three different epochs ($M_{\rm tot}=1$, $2$, and $4\,\msun$) for the aligned models with different $\mu_0$ against the enclosed mass $M_{\rm enc}$, in which $J_{\rm enc}$ and $M_{\rm enc}$ are estimated for a radius $r$ to be
\begin{equation}
J_{\rm enc} (r) = J_{\rm b} + \int_0^{r} \rho \, r_{\rm c} v_\phi \, dv,
\end{equation}
where $r_{\rm c}$, $v_\phi$, and $J_{\rm b}$ are the distance from the rotation axis,  the velocity in the azimuthal direction, and the angular momentum that has already fallen onto the sink, respectively, and  
\begin{equation}
M_{\rm enc} (r) = M_{\rm tot} + \int_0^{r} \rho \, dv,
\end{equation}
in which $M_{\rm tot}$ is the total mass fallen onto the sink. 
The panel indicates that angular momenta for these models are smaller than the initial value near the center.
Note that the enclosed mass $M_{\rm enc}$ is an increasing function of radius. 
For clarity, we plot the angular momentum normalized by the initial value in Figure~\ref{fig:compare_mu_profile} right panel. 
From these panels, we can confirm that the angular momentum transfer occurs near the center of the cloud \citep[for details, see][]{machida11}.
The angular momentum near the center ($ M_{\rm enc} \lesssim 5 \msun$) is transported into the outer layer of the infalling envelope. 
For the models with $\mu_0=1.5$ and 2.0, over 90\% of the initial angular momentum is transported into the infalling envelope just outside the sink when the binary mass is $M_{\rm tot}=2\,\msun$.
On the other hand, at the same epoch, about 60\% of the initial angular momentum is transported near the sink for the model with $\mu_0=3.0$. 
Thus, the efficiency of angular momentum transport significantly depends on the initial magnetic field strength.

\begin{figure*}
\includegraphics[width=0.8\columnwidth]{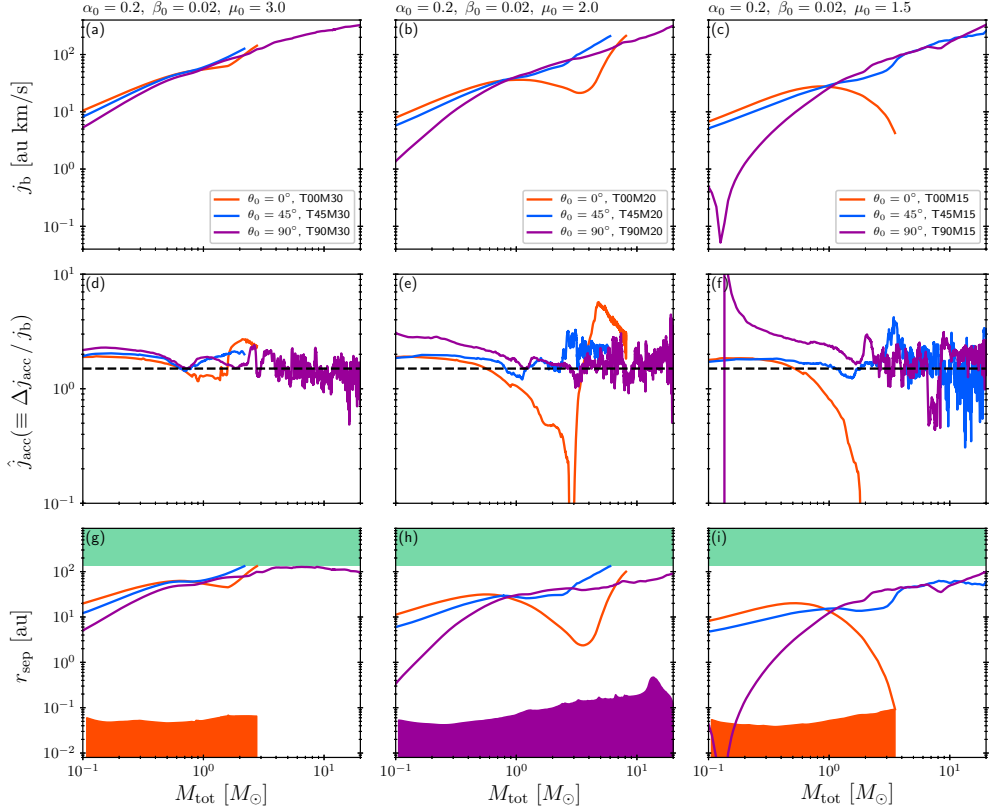}
\caption{
Same as in Fig.~\ref{fig:compare_mu}.
Models with different $\theta_0$ but with the same $\mu_0$ are plotted in each panel. 
Parameters $\alpha_0$, $\beta_0$, and $\mu_0$ are given in each top panel.
}
\label{fig:compare_theta}
\end{figure*}

\subsubsection{Dependence on the Angle between Rotation Axis and Magnetic Field}
\label{sec:parameter_theta}

As described in \S\ref{sec:paramag}, the efficiency of angular momentum transport also depends on the angle between the magnetic field and the rotation axis. 
\citet{Hirano2020} showed that the configuration of the magnetic field significantly changes over time for both the aligned ($\theta_0=0^\circ$) and misaligned ($\theta_0 \ne 0^\circ$) models.
They also pointed out that  the efficiency of  magnetic braking strongly depends on the configuration of the magnetic field (see also \citealt{hennebelle09, joos12, tsukamoto18}).
We briefly explain the results shown in \citet{Hirano2020}.
For the aligned model,  the magnetic field lines are nearly parallel in the early main accretion phase, while they have a fan-shaped configuration in the late accretion phase.
Magnetic braking is more efficient in the fan-shaped configuration than in the parallel configuration. 
We also consider the magnetic field lines perpendicular to the rotation axis (perpendicular case, $\theta_0=90^\circ$).
Magnetic braking is more efficient in the perpendicular case than in the (aligned) parallel case, but less efficient than in the (aligned) fan-shaped case. 
Thus, in addition to the initial configuration of the magnetic field, the time evolution of the configuration is also important for evaluating the efficiency of magnetic braking.

$j_{\rm b}$, $\hat{j}_{\rm acc}$, and $r_{\rm sep}$ for the models with different $\theta_0$ are plotted in Figure~\ref{fig:compare_theta}. 
There is no significant difference in these quantities between  the models when the initial magnetic field is relatively weak ($\mu_0=3.0$; Figs.~\ref{fig:compare_theta}{\it a}, {\it d}, and {\it g}).
However, only the model with $\theta_0=90^\circ$ did not reach the upper limit of binary separation. 
For this model, the ratio $\hat{j}_{\rm acc}$ oscillates around the critical value $\hat{j}_{\rm acc} \sim  \hat{j}_{\rm cri}$ for $M_{\rm tot} \gtrsim 2\msun$  (Fig.~\ref{fig:compare_theta}{\it d}), which indicates that $r_{\rm sep}$ does not monotonically increase in the late main accretion phase.
As a result, $r_{\rm sep}$ is marginally within the upper limit ($r_{\rm sep} < 2\, r_{\rm sink}$) by the end of the simulation.

Among the models with $\mu_0=2.0$ (Figs.~\ref{fig:compare_theta}{\it b}, {\it e}, and {\it h}), only the model with $\theta_0=90^\circ$  became a close binary system.
Note that  we stopped the calculation for the model with $\theta_0 = 0^\circ$, even though the binary separation reached neither the lower or upper limit.
This is because the calculation timestep becomes very short due to the emergence of a high-velocity outflow, and the calculation was not completed within the limited time.
Thus, for this model, we could not integrate the calculation until the total stellar mass reached $M_{\rm tot}=20\msun$. 

We can see a significant difference between the aligned ($\theta_0=0^\circ$) and misaligned ($\theta_0 \ne 0^\circ$) models when the initial magnetic field strength is as strong as $\mu_0=1.5$ (Figs.~\ref{fig:compare_theta}{\it c}, {\it f}, and {\it i}). 
For these models, the binary separation in the aligned model $\theta_0=0^\circ$ is the smallest and finally reaches the lower limit of twice the stellar radius.
On the other hand, the models with $\theta=45$ and $90^\circ$ reach neither the lower or upper limit
\footnote{
As described in \S\ref{sec:paramag},  for model T90M15, we consider the evolution of the binary system only for $M_{\rm tot}>0.2\msun$.
In addition, only for this model, we confirmed a sign inversion of the angular momentum in an early phase ($7\times10^{-4}\msun \lesssim M_{\rm tot} \lesssim 0.13\msun$) during which the angular momentum becomes negative with a mass of $\sim0.1\msun$. 
A sign inversion means that the sign of the angular momentum changed from plus to minus or minus to plus (for details, see \citealt{mouschovias79,machida2020c}).
}.
Thus, the close binary system is maintained by the end of the simulation for these models.
The specific angular momentum and binary separation for these models ($\theta=45$ and $90^\circ$) gradually increase (Fig.~\ref{fig:compare_theta}{\it c} and {\it i}), while  $\hat{j}_{\rm acc}$ oscillates around the critical value $\hat{j}_{\rm acc} \sim \hat{j}_{\rm cri}$ (Fig.~\ref{fig:compare_theta}{\it f}). 

In addition, in Figure~\ref{fig:compare_theta}{\it i}, the binary separation for the model with $\theta= 90^\circ$ is much smaller than that for the model with $\theta_0=0^\circ$  in the early phase of $0.2\msun < M_{\rm tot}<1\msun$, while the magnitude relation of the binary separation between $\theta_0=0$ and $90^\circ$ models is inverted in the late phase of $M_{\rm tot}>1\msun$. 
This indicates that the configuration of the magnetic field changes over time.
Especially,  magnetic braking is more efficient in the late phase  than in the early phase for the aligned case, as described in \citet{Hirano2020}.
For the aligned case ($\theta_0=0^\circ$), the efficiency of magnetic braking becomes gradually effective as time proceeds because the configuration of the magnetic field changes from the aligned uniform to the aligned fan-shaped configuration, as described above  (for details, see Fig.~1 of \citealt{Hirano2020}).  
On the other hand, for the misaligned ($\theta_0=45^\circ$) or perpendicular ($\theta_0=90^\circ$) cases, the efficiency of magnetic braking is larger than for the aligned uniform case but smaller than for the aligned fan-shaped case independent of the evolutionary stage \citep{Hirano2020}. 
As a result, the binary separation in the aligned model is smaller than that in the misaligned model  in the late accretion phase, because the efficiency of  magnetic braking is most effective in the aligned model (or the aligned fan-shaped configuration).

\begin{figure*}
\includegraphics[width=0.8\columnwidth]{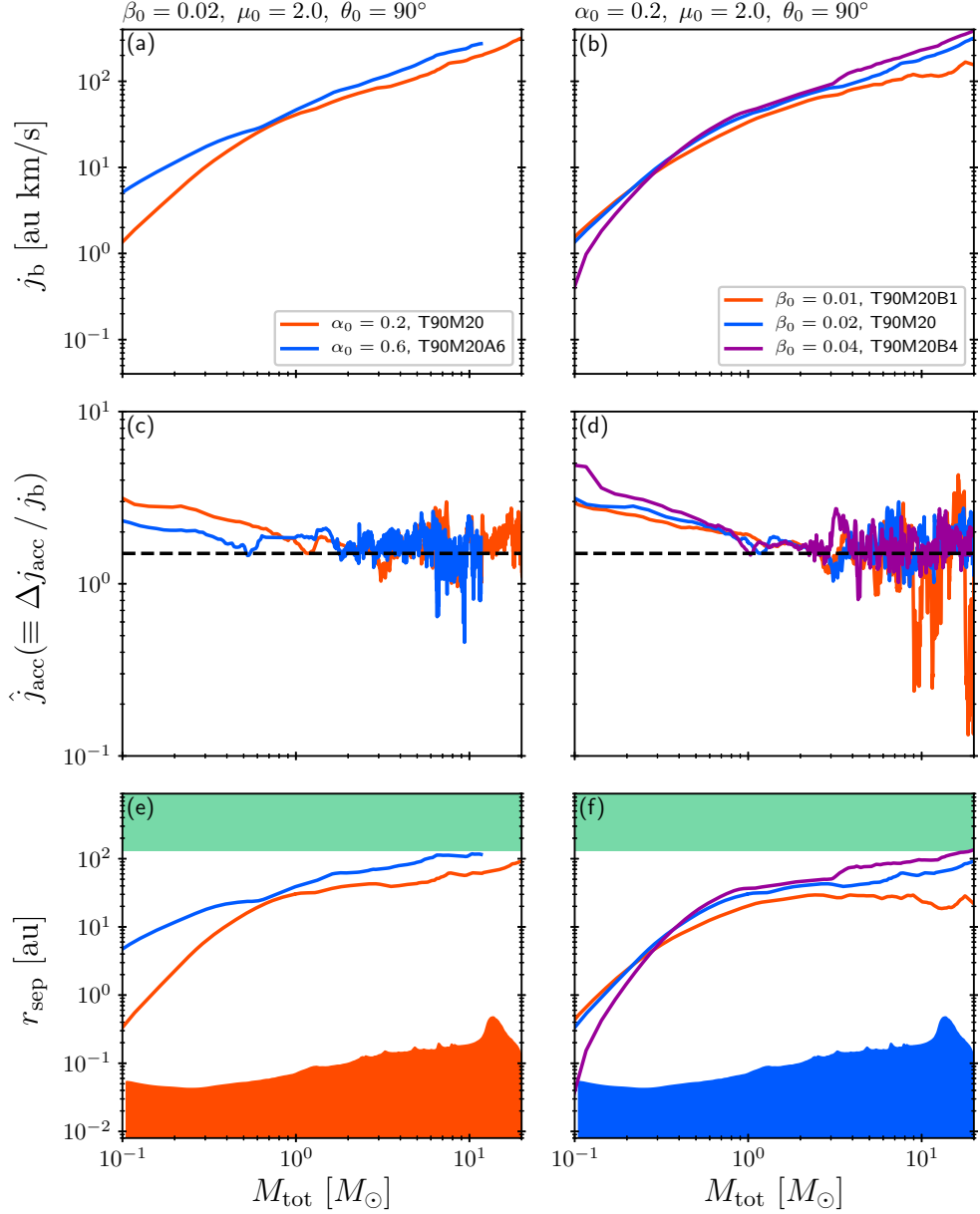}
\caption{
Binary specific angular momentum $j_{\rm b}$ (top panels), accreting angular momentum $\hat{j}_{\rm acc}$ (middle panels), and binary separation $r_{\rm sep}$ (bottom panels) against total stellar mass $M_{\rm tot}$.
Models with different $\alpha_0$ but with the same $\beta_0$,  $\mu_0$, and $\theta_0$ are plotted in the left panels.
Models with different $\beta_0$ but with the same $\alpha_0$,  $\mu_0$, and $\theta_0$ are plotted in the right panels.
The broken line in the middle panels indicates the critical angular momentum $\hat{j}_{\rm cri}$.
In the bottom panels, the boundary between the lower colored and white areas corresponds to twice the stellar radius, in which the color corresponds to the model with the same line color.
The green area in the bottom panels is the region outside the sink.  
}
\label{fig:compare_beta}
\end{figure*}

\subsubsection{Dependence on parameters $\alpha_0$ and $\beta_0$}
In this subsection, we investigate the dependence of the binary separation on $\alpha_0$ and $\beta_0$. 
Note that  the models with $\theta_0=0$ and 45$^\circ$ tend to become a wide binary system, as described above.
These models are not adequate for investigating the long-term evolution of the binary separation, because the calculation was stopped just after the binary separation reached the upper limit. 
Thus, in this subsection, we consider only the models with $\theta_0=90^\circ$ to investigate the parameter dependence of $\alpha_0$ and $\beta_0$  in the long-term evolution.

Figure~\ref{fig:compare_beta} left panels show $j_{\rm b}$ (Fig.~\ref{fig:compare_beta}{\it a}), $\hat{j}_{\rm acc}$ (Fig.~\ref{fig:compare_beta}{\it c}), and $r_{\rm sep}$ (Fig.~\ref{fig:compare_beta}{\it e}) against the total stellar mass for the models with different $\alpha_0$ but with the same $\beta_0$, $\mu_0$, and $\theta_0$. 
The mass accretion rate is controlled by the parameter $\alpha_0$, and is large for small $\alpha_0$ \citep{Matsushita2017,Machida2020}
\footnote{
The dependence of the mass accretion rate on the parameters $\mu_0$ and $\theta_0$ has been investigated in \citet{Machida2020} and \citet{Hirano2020}. 
The dependence on the parameter $\beta_0$ has not been investigated.
However, it is expected that  the parameter $\beta_0$ does not significantly affect the mass accretion rate, because the rotational energy is lower than the magnetic energy in star-forming cores, which does not affect the mass accretion rate \citep{Machida2020}.
}. 
In addition, the configuration of the magnetic field should differ with the parameter $\alpha_0$, because the geometry of the infalling envelope differs with different $\alpha_0$ \citep{Hirano2020,machida20}.
Nevertheless, there is no significant difference between models with $\alpha_0=0.2$ and 0.6.
$j_{\rm b}$ is larger in the model with $\alpha_0=0.6$ than in the model with $\alpha_0=0.2$, while the difference between them at the end of the simulation is not significant.
Thus, the initial cloud stability (or parameter $\alpha_0$) does not significantly affect the determination of binary separation.

$j_{\rm b}$, $\hat{j}_{\rm acc}$, and $r_{\rm sep}$ with different $\beta_0$ are plotted against the total mass in Figure~\ref{fig:compare_beta} right panels, in which the parameters $\alpha_0$, $\mu_0$, and $\theta_0$ are fixed. 
Figure~\ref{fig:compare_beta}{\it f}~ indicates that a small $\beta_0$ results in a narrow  $r_{\rm sep}$. 
However, as seen in Figure~\ref{fig:compare_beta}{\it d},  $\hat{j}_{\rm acc}$ strongly oscillates around $\hat{j}_{\rm acc} \sim \hat{j}_{\rm cri}$ in each model.
Thus, it is difficult to predict the final $r_{\rm sep}$ from the initial rotational energy of the cloud, because the angular  momentum transfer in the infalling envelope is very complicated. 

\subsection{Outcome of Binary Separation}
\begin{figure*}
\includegraphics[width=0.9\columnwidth]{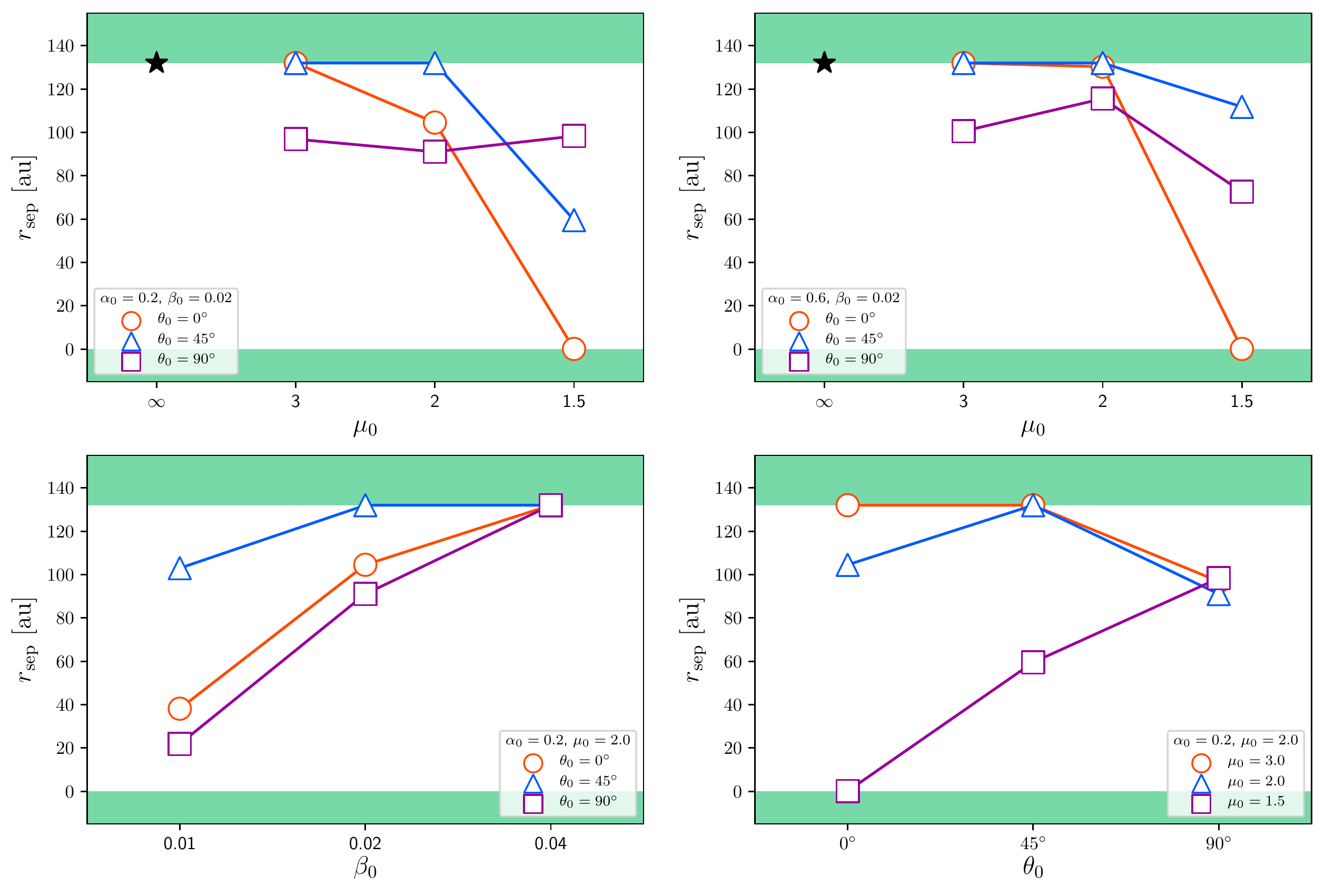}
\caption{
Binary separation $r_{\rm sep}$ at the end of the simulation against the parameters  $\mu_0$ (top panels), $\beta_0$ (bottom left), and $\theta_0$ (bottom right). 
In the top panels, the models with $\alpha_0=0.2$ (left) and $0.6$ (right) are plotted, in which an unmagnetized model is plotted by the black star symbol.   
The upper limit corresponds to twice the sink radius, while the lower limit is set to be $r_{\rm sep}=0$ in each panel.
The model parameters are described in each panel. 
}
\label{fig:final_separation}
\end{figure*}

Figure~\ref{fig:final_separation} shows the binary separation $r_{\rm sep}$ at the end of the simulation against the parameters $\mu_0$, $\beta_0$, and $\theta_0$.
The models with different $\alpha_0$ are plotted in the top panels of Figure~\ref{fig:final_separation}. 
Although the dependence of $r_{\rm sep}$ on the parameter $\alpha_0$ is not clear,  the dependence on the parameter $\mu_0$ is clear.
The binary separation $r_{\rm sep}$ becomes narrow as the magnetic field becomes strong except for the model with $\theta_0=90^\circ$. 
In the range $\mu_0=1.5$--$3$, the final binary separation is not significantly affected by the magnetic field strength for the models with $\theta_0=90^\circ$. 
These panels also indicate that, depending on the parameters $\alpha_0$ and $\theta_0$,  a close binary system can form when the initial cloud has a strong magnetic field ($\mu_0= 1.5$--$3$).

The dependence of the binary separation on $\beta_0$ is obvious (Fig.~\ref{fig:final_separation} bottom left).
The binary separation shortens as $\beta_0$ decreases. 
It is natural that the initial rotational energy influences the binary separation. 
The initial angle $\theta_0$ between the magnetic field and the rotation axis also affects the determination of $r_{\rm sep}$. 
However, we cannot see a clear trend of $\theta_0$ dependence in the bottom right panel of Figure~\ref{fig:final_separation}. 
The parameter $\theta_0$ is related to the efficiency of magnetic braking  and the strength of outflow, both of which determine the angular momentum fallen onto the sink and the binary separation. 
The configuration of the magnetic field is very complicated and changes over time in the misaligned models, indicating that the angular momentum introduced into the central region changes from time to time (see \S\ref{sec:oscillation}). 
As a result, it is difficult to find  the dependence of $r_{\rm sep}$ on  $\theta_0$.

Although the parameter dependence of $\alpha_0$ and $\theta_0$ on the binary separation is not clear, we can state that, in addition to the initial rotational energy,  the magnetic field strength is an important factor for determining the binary separation.

\section{Discussion}
\subsection{Oscillation in Accreting Angular Momentum} 
\label{sec:oscillation}

In Figures~\ref{fig:compare_mu}{\it e}, \ref{fig:compare_mu}{\it f}, \ref{fig:compare_theta}{\it d}, \ref{fig:compare_theta}{\it e}, \ref{fig:compare_theta}{\it f}, \ref{fig:compare_beta}{\it c}, and \ref{fig:compare_beta}{\it d}, we can see the oscillation in $\hat{j}_{\rm acc}$ in the misaligned models ($\theta_0=45$ and $90^\circ$). 
On the other hand, the oscillation of $\hat{j}_{\rm acc}$ does not occur in the aligned models  ($\theta_0=0^\circ$). 
In this subsection, we discuss the cause of the oscillation.

\begin{figure*}
\includegraphics[width=0.8\columnwidth]{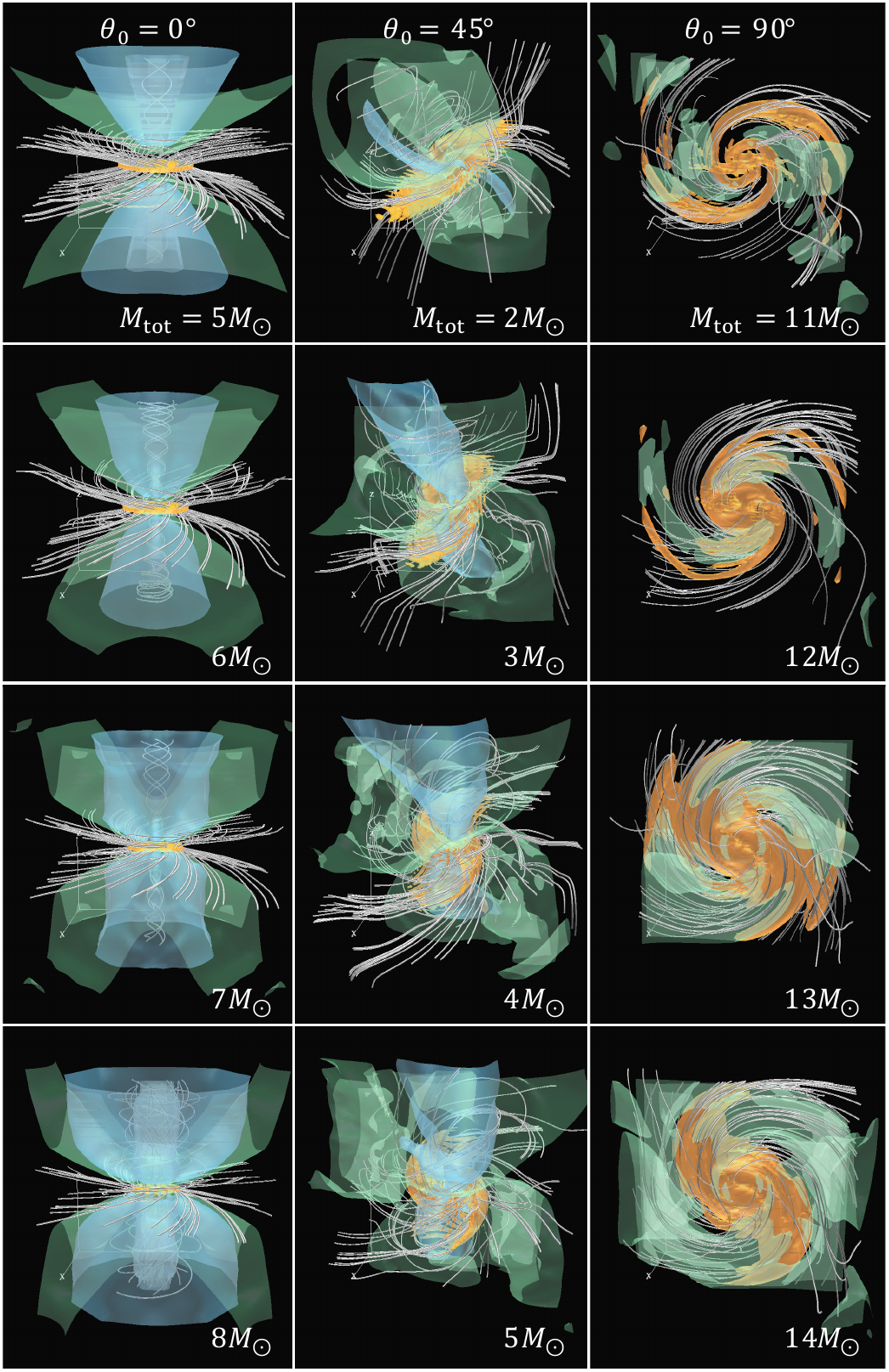}
\caption{
Time sequence of models T00M20 ($\theta_0=0^\circ$; left), T45M20 ($\theta_0=45^\circ$; middle), and T90M20 ($\theta_0=90^\circ$; right).
The total mass $M_{\rm tot}$ is given in each panel.
A three-dimensional structure is plotted in each panel. 
The white lines correspond to magnetic field lines.
The yellow surface is the iso-density surface of $8.9\times10^7$\,cm$^{-3}$ corresponding to the pseudo disk.
The outflow is represented by the green and blue surfaces, which correspond to the iso-velocity surfaces of $v_r=0.13$ (green) and $2.6$\,km\,s$^{-1}$ (blue), respectively. 
The box size is $\sim 10^3$\,au in all panels.
}
\label{fig:multi_bfield}
\end{figure*}

In Figures~\ref{fig:compare_mu}{\it e} and \ref{fig:compare_mu}{\it f}, for the models with $\theta_0=45^\circ$ (T45M20) and $90^\circ$ (T90M20), the oscillation occurs in the range $M_{\rm tot} > 2\msun$. 
Thus, we plot three-dimensional structures after the total mass reaches $M_{\rm tot}>2\msun$ for models T00M20, T45M20, and T90M20  in Figure~\ref{fig:multi_bfield}, in order to investigate the condition under which the oscillation occurs.

For the aligned model (left column), the structures of the pseudo disk, outflow, and magnetic field do not significantly change over  time. 
Although the outflow opening angle gradually widens with time, we cannot see a noticeable change in the configuration of the magnetic field and the disk-like structure for this model.
As described in \S\ref{sec:results}, the accreting angular momentum $\hat{j}_{\rm acc}$ does not show a strong oscillation for the aligned models. 
Thus, it is considered that the angular momentum is steadily removed by magnetic braking and outflow for this model.

On the other hand, in the misaligned models T45M20 (middle panels) and T90M20 (right panels), the structures change violently. 
In the middle panels, we can confirm that the outflow structure and  the magnetic configuration significantly change over time. 
In the right panels, the structure of the high-density region  (yellow surface) corresponding to the iso-density surface of $8.9\times10^7$\,$\cm$ changes over time. 
The time variation of the structure  seen in the misaligned models (middle and right panels) is considered to be related to the efficiency of the angular momentum transfer due to magnetic braking.

The pseudo disk (or infalling envelope) is mainly supported by the Lorentz force and is partly supported by the rotation (or centrifugal force). 
When a strong magnetic field and a dense envelope exist around the sink, the angular momentum near the sink would be effectively removed by magnetic braking \citep{Hennebelle2009}. 
Just after the removal of the angular momentum, a part of the gas around the sink would rapidly falls onto the sink.  
Then, the pseudo disk (or envelope) becomes less massive and magnetic braking becomes less effective, which results in the accumulation of the gas around the sink.
The outflow structure and magnetic configuration would also change according to the change in the density distribution around the sink.
The density distribution and magnetic field configuration change with time, and they significantly affect the efficiency of  magnetic braking \citep{mouschovias79}. 
Thus, it is expected that the oscillation of $\hat{j}_{\rm acc}$ can be attributed to the rapid structural change around the sink.

\subsection{Binary Separation and Spatial Resolution in Simulation}
\label{sec:resoluton}

The binary separation is an important factor determining the properties of binary systems. 
Especially, the formation of close massive binaries should clarify the origin of objects detected in gravitational waves. 
Observations indicate a high existence probability of close massive binary systems composed of equal mass stars.  
However, it is difficult to form such a system because massive stars (or a massive binary) should form in a large (massive) cloud with a large angular momentum.
It is expected that a large angular momentum would result in the formation of a wide binary system \citep{bodenheimer11}. 
Thus, angular momentum must be removed from a star-forming cloud to form close binary systems.
Magnetic braking is a promising candidate for the removal of  angular momentum. 

This study focuses on angular momentum removal in massive star-forming clouds by  magnetic braking. 
As described in \S\ref{sec:methods}, we deliberately did not spatially resolve the binary system, in order to investigate the angular momentum falling onto the central region over a long duration. 
In other words, the numerical simulations were only used  to precisely estimate the configuration and strength of  the magnetic field and the efficiency of magnetic braking without resolving the binary system. 
Thus, the spatial resolution of the simulation is not very high, as described in \S\ref{sec:mhdsimulation}.

Our purpose is to qualitatively investigate whether magnetic braking affects the formation of close binary systems. Our  aim is not to quantitatively determine the binary separation. 
However, it  would be of benefit to discuss the resolution dependence on the results.  
To investigate this, we calculated three additional models with the same parameters $\mu_0$, $\beta_0$, $\alpha_0$, and $\theta_0$ but with different sink threshold densities and radii (or different spatial resolutions). 

\label{sec:resolution}
\begin{figure*}
\includegraphics[width=\columnwidth]{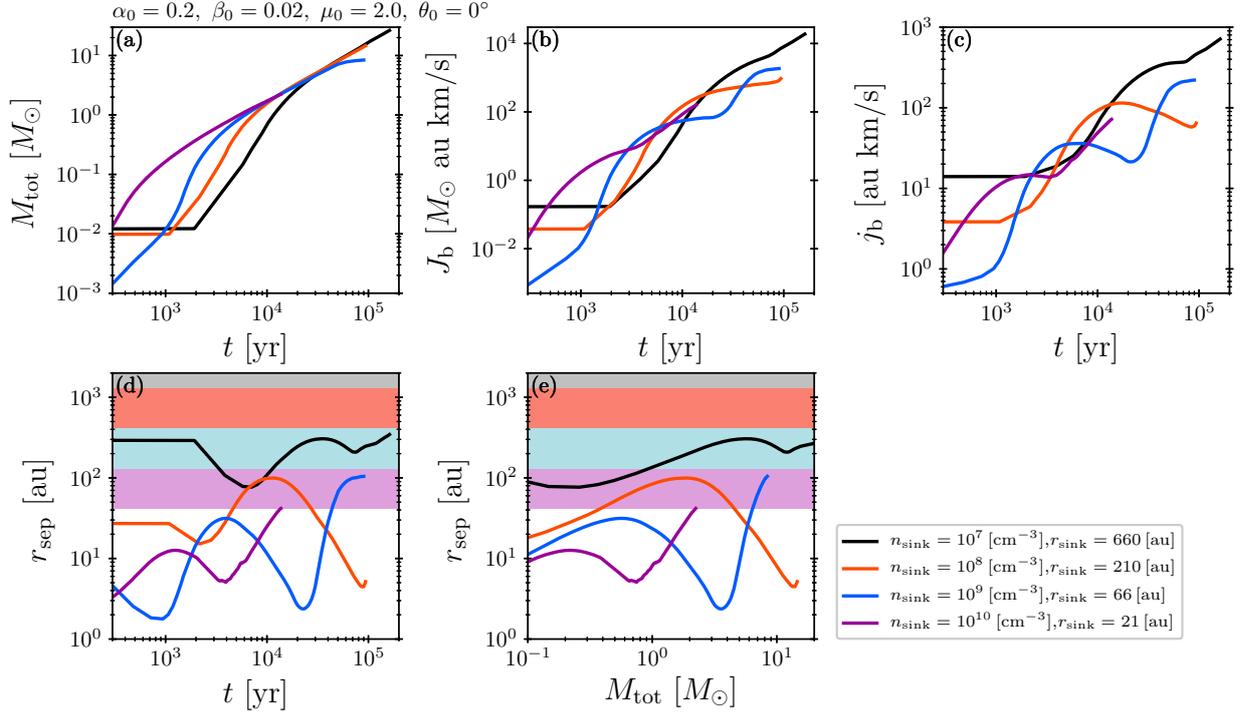}
\caption{
Total mass $M_{\rm tot}$,  total angular momentum $J_{\rm b}$, and specific angular momentum $j_{\rm b}$ calculated from simulations with different sink conditions, plotted against the elapsed time from the formation of the binary $t$ in panels (a), (b), and (c), respectively.
The binary separation $r_{\rm sep}$ estimated from the simulations shown in panels (a), (b), and (c) is plotted against the elapsed time $t$ and total mass $M_{\rm tot}$ in panels (d) and (e).
The model parameters are the same ($\mu_0=2.0$, $\beta_0=0.02$, $\alpha_0=0.2$, and $\theta_0=0^\circ$).
The adopted sink threshold densities $n_{\rm sink}$ and sink radii $r_{\rm sink}$ are described in the lower right region.
The colored area in panels (d) and (e) is the upper limit of the binary separation (or the region outside the sink radius), in which the color of each area corresponds to that of each solid line.
}
\label{fig:convergence}
\end{figure*}

\begin{figure*}
\centering
\includegraphics[width=0.9\columnwidth]{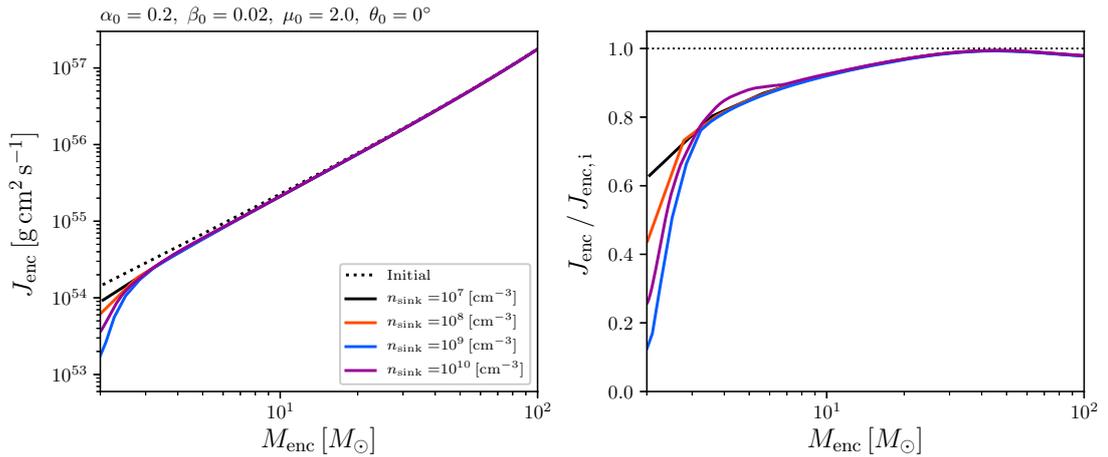}
\caption{
Distribution of enclosed angular momentum (left) and enclosed angular momentum divided by the initial value (right) against the enclosed mass for the models with different sink conditions when the binary mass reaches $M_{\rm tot}=2\,\msun$. 
The black dotted line in both panels is the initial distribution of the enclosed angular momentum. 
The parameters $\alpha_0$, $\beta_0$, $\mu_0$, and $\theta_0$ are given in the upper part of the left panel.
}
\label{fig:convergence2}
\end{figure*}

The total mass $M_{\rm tot}$, angular momentum $J_{\rm b}$, and specific angular momentum $j_{\rm b}$ calculated with different sink criteria are plotted in Figures~\ref{fig:convergence}{\it a}--{\it c}, in which we only changed the sink threshold densities and corresponding sink radii.
Note that the sink radii are related to the sink threshold densities through the Jeans length as  $r_{\rm sink}=\lambda_{\rm J}/2$ (see also eq.~[\ref{eq:r_sink}]).
It is difficult to confirm the differences in the total mass and angular momentum because they do not follow a monotonically increasing function. 
However, it seems that there is no obvious significant difference between the calculations with different sink criteria. 

To further investigate the effect of the sink, the enclosed angular momentum for the models with different sink conditions is plotted against the enclosed mass in Figure~\ref{fig:convergence2}, in which the binary mass are the same ($M_{\rm tot}=2\msun$) in all the models.  
As seen in the figure, the difference in the angular momenta appears only around the central region ($M_{\rm enc}\sim 2$--$3\msun$). 
The difference is caused by different efficiencies of the angular momentum transfer due to magnetic braking. 
However, we can see a convergence of the angular momentum around $M_{\rm enc}\simeq3\msun$. 
Note that, in the right panel of Figure~\ref{fig:convergence2}, although $J_{\rm enc}/J_{\rm enc,i}$ for the model with $n_{\rm sink}=10^{10}\cm$ deviates from those for other models, the deviation is not significant. 
Thus, although different sink conditions (or properties) influence the efficiency of the magnetic braking near the sink, it does not significantly affect the angular momentum distribution at a large scale. 
We concluded that although further high-spatial-resolution simulations are necessary, we can roughly estimate the angular momentum falling onto the binary system with a sink.

The binary separation against the elapsed time and total mass estimated from the simulations with different sink criteria is plotted in Figures~\ref{fig:convergence}{\it d} and \ref{fig:convergence}{\it e}. 
In Figure~\ref{fig:convergence}{\it d}, the binary separations are plotted since 300\,yr after binary formation to confirm the early evolution of the binary separation.
Note that since we output the data for these models every $\sim 10-100$\,yr, it is difficult to plot the binary separation for $t \lesssim 300$\,yr.
We did not initially expect convergence of the binary separation. 
However, the figures show that the binary separations are within the range $2<r_{\rm sep}<100\,{\rm au}$ except for the coarsest model having $n_{\rm thr}=10^7\cm$ and $r_{\rm sink}=660$\,au.
It seems that the binary separation tends to shrink as the spatial resolution becomes high (or the sink radius decreases) in these calculations.
On the other hand,  the  binary separations of these models are considerably smaller than that of the unmagnetized model (see the black line in Fig.~\ref{fig:compare_mu}{\it g}), which indicates that  magnetic braking is important for the formation of close binary systems.
We would require huge computational resources  to precisely determine the final binary separation until the main accretion phase ends. 
Such calculations may be possible in future studies.

\subsection{Other Mechanisms of Angular Momentum Transport}
\label{sec:other}
In this subsection, we describe other mechanisms of angular momentum transport. 
Although this study focused on angular momentum transport due to  magnetic braking on a large scale, angular momentum is also transported by other effects on different scales. 
The binary orbital motion causes a gravitational torque that transports angular momentum around the circumbinary disk outward. 
As seen in the observations, in addition to the low-velocity outflow driven by the circum-binary disk \citep{machida09}, high-velocity jets   appear from each protostar \citep{hara20}. 
Both outflows and jets can transfer angular momentum \citep{Saiki2020}. 
If these effects are included, we can only calculate the binary evolution for $\sim400$\,yr as seen in \citet{Saiki2020}. 

We also ignored the effect of turbulence. 
\citet{Joos2013} and \citet{Seifried2013} pointed out that turbulence can alleviate the magnetic braking, which may widen the binary separation. 
However, \citet{Lund2018} pointed out the binary separation tends to decrease in a turbulent cloud because gas with small specific angular momentum can be unevenly present in such a cloud. 
In addition, the angular momentum would be transferred by the Reynolds stress if strong turbulence is present around the binary system \citep{abel2002}.
Thus, since turbulence seems to play some different roles in the binary formation process, we cannot clearly explain the effect of turbulence on the binary evolution.

Since we ignored some mechanisms of angular momentum transport at small scales, our results would give an upper limit for binary separation. 
Our future study will focus on some of other mechanisms of angular momentum transport.

\subsection{Comparison with Lund \& Bonnell}

This study is motivated by \citet{Lund2018}. 
The binary separations were analytically estimated in \citet{Lund2018}, while they were numerically derived in this study. 
In both studies, we can confirm the same trend that the binary separation decreases as the magnetic field becomes strong.

We roughly compare our results with theirs. 
We adopted a magnetic field strength in the range $0.86$--$5.2\,\mu$G for magnetized models, as listed in Table~\ref{tab:initial_condition}. 
Since \citet{Lund2018} only assumed the aligned case (magnetic field parallel to the rotation axis), we select the results with $\theta_0=0^\circ$ (aligned cases). 
As seen in Figures~\ref{fig:compare_mu} and \ref{fig:final_separation}, the binary separations are in the range $\sim0.1$--$100$\,au in our study. 
On the other hand, in \citet{Lund2018}, the binary separation is about 10\,au (about $5$\,au) when a magnetic field strength of $1\,\mu$G ($10\,\mu$G) is adopted for  their fiducial model (see Fig.~3{\it c} of \citealt{Lund2018}).
Thus, although the numerical settings differ considerably, the binary separations derived in this study do not greatly  contradict  those derived in \citet{Lund2018}.
It should be noted that since there are many parameters to determine the initial conditions and numerical settings, it is difficult to fairly compare the results of both studies. 
Actually, in addition to the magnetic field strength, other factors such as the initial cloud mass, rotation rate, cloud stability, and initial angle between the initial magnetic field and rotation axis can change the binary separation.
However, \citet{Lund2018} and this study indicate that the magnetic field is  a crucial ingredient for the formation of high-mass close binary systems.

\section{Conclusions}

Using 3D MHD simulations, we tracked the evolution of rotating magnetized clouds with four different parameters $\mu_0$, $\beta_0$, $\alpha_0$, and $\theta_0$ corresponding to the magnetic field strength, degree of rotation,  initial cloud stability, and angle between the initial magnetic field and rotation axis, respectively.
In the simulations, a large sink radius ($\sim66$\,au) was adopted for the fiducial models to realize a long-term evolution ($\sim10^5$\,yr), in which the mass within the sink was treated as a point mass.
Then, we investigated the mass and angular momentum falling onto the central region (or sink)  in order to estimate how much angular momentum is removed before a parcel of gas reaches the central protostars (or proto-binary system).
Finally, assuming an equal mass binary system with a circular orbit,  we  analytically estimated the binary separation to determine whether close binary systems can be formed through the effect of magnetic braking.  
We obtained the following results.

\begin{itemize}
\item 
The magnetic field of a star-forming cloud is the most important factor in determining whether a close binary system is formed. 
In an unmagnetized cloud, the binary separation (or total angular momentum falling onto the sink) continues to increase and a wide binary with a separation of $>100$\,au forms.
The binary separation decreases when the star-forming cloud is magnetized.
Thus, close binary systems can form in magnetized clouds.
\item 
For the aligned models ($\theta_0=0^\circ$),  the binary separation increases in the early accretion phase.
Then, after the binary separation temporally decreases, it increases again in the late accretion phase. 
The temporal decrease is considered to be caused by a change of the configuration of the magnetic field. 
Magnetic braking is more efficient  in the fan-shaped configuration than in the parallel configuration when the initial magnetic vector is parallel to the initial angular momentum vector. 
Thus, the angular momentum is efficiently removed with the fan-shaped configuration, which is realized in the late accretion phase. 
Meanwhile, only a large angular momentum can accrete onto the system in the late accretion phase.
A powerful outflow appears in the aligned models.
The outflow opening angle widens with time, and the accretion of gas with small angular momentum distributed along the rotation  axis is interrupted by the outflow. 
As a result, the binary separation (and accreted angular momentum) significantly changes over time for the aligned models. 
\item 
For the misaligned models ($\theta_0\ne0^\circ$),  the accreting angular momentum and  binary separation continue to increase in the early accretion phase.
However, in the late accretion phase, the specific angular momentum of accreting matter is comparable to or smaller than that of the binary system.
Thus, the increase rate of the binary separation is small or the binary separation shrinks.
As a result, a close binary forms.
\item
When the initial cloud has a strong magnetic field, the angular momentum is excessively removed during the main accretion phase. 
Thus, the binary separation is smaller than the stellar radius, which means that a merger between protostars occurs. 
This may indicate that close binary formation occurs in a limited parameter space.
\item
Outflow is more apparent in  the aligned models ($\theta_0=0^\circ$) than in the misaligned models ($\theta_0\ne0^\circ$), which is consistent with previous studies.
Thus, the outflow does not significantly interrupt the accretion of the gas with small angular momentum for the misaligned models ($\theta_0\ne0^\circ$). 
Therefore, the protostellar outflow and the initial angle between the magnetic field and rotation axis  are also  important factors for determining the angular momentum falling onto the binary system and the resultant binary separation.

\end{itemize}

We conclude that the strength and configuration of the initial magnetic field in a star-forming cloud are important for investigating close binary formation and a cloud with a strong magnetic field that is not parallel to the rotation axis is a favorable environment for the formation of close binary systems.
Our study supports the claim  that magnetic braking is crucial for close binary formation, as pointed out by \citet{Lund2018}. 

\section*{Acknowledgements}
This research used the computational resources of the HPCI system provided by the Cyber Science Center at Tohoku University, the Cybermedia Center at Osaka University, and the Earth Simulator at JAMSTEC through the HPCI System Research Project (Project ID: hp180001, hp190035, hp200004, hp210004).
The simulations reported in this paper were also performed by 2019 and 2020 Koubo Kadai on the Earth Simulator (NEC SX-ACE) at JAMSTEC. 
The present study was supported  by JSPS KAKENHI Grants (JP17H02869, JP17H06360, JP17K05387, JP17KK0096, JP21K03617, JP	21H00046: MNM, JP19H01934: TH).

\section*{DATA AVAILABILITY}
The data underlying this article are available in the article.



\bibliographystyle{mnras}
\bibliography{ref_article} 





\bsp	
\label{lastpage}
\end{document}